\documentclass[preprintnumbers,article,amsmath,amssymb,floatfix,10pt,prd,onecolumn,
superscriptaddress,nofootinbib]{revtex4}
\usepackage[colorlinks=true, pdfstartview=FitV, linkcolor=blue, citecolor=red, urlcolor=magenta]{hyperref}
\usepackage{bbm}
\usepackage{amsfonts}
\usepackage{mathrsfs}
\usepackage{latexsym}
\usepackage{epsfig}
\usepackage{epstopdf}
\usepackage{epstopdf}
\usepackage{graphicx}
\usepackage{amssymb}
\usepackage{amsmath}
\usepackage{dcolumn}
\usepackage{bm}
\usepackage{float}
\usepackage{color}
\usepackage{comment}
\usepackage{xcolor}
\begin{document}

\title{\bf Spherically Symmetric Wormhole Solutions admitting Karmarkar Condition}

\author{M. Zeeshan Gul}
\email{mzeeshangul.math@gmail.com}\affiliation{Department of Mathematics and Statistics, The University of Lahore,\\
1-KM Defence Road Lahore-54000, Pakistan.}

\author{M. Sharif}
\email{msharif.math@pu.edu.pk}\affiliation{Department of Mathematics and Statistics, The University of Lahore,\\
1-KM Defence Road Lahore-54000, Pakistan.}

\begin{abstract}
This paper investigated the viable traversable wormhole solutions
through Karmarkar condition in the context of $f(\mathcal{G},T)$
theory. A static spherical spacetime with anisotropic matter
configuration is used to study the wormhole geometry. Karmarkar
condition is used to develop a viable shape function for a static
wormhole structure. A wormhole geometry is constructed using the
resulting shape function that satisfies all the required conditions
and connects the asymptotically flat regions of the spacetime. To
assess the viability of traversable wormhole geometries, the energy
conditions are analyzed by various models of this theory. Further,
their stable state is investigated through sound speed and adiabatic
index. This investigation demonstrates the presence of viable
traversable wormhole solutions in the modified
theory.\\\\
\textbf{Keywords}: Wormhole; $f(\mathcal{G},T)$ theory; Karmarkar
condition; Stability analysis.
\end{abstract}

\maketitle

\date{\today}

%%%%%%%%%%%%%%%%%%%%%%%%%%%%%%%%%%%%%%%%%%%%%%%%%%%%%%%%%%%%%%%%%%%%%%%%
%%%%%%%%%%%%%%%        Introduction        %%%%%%%%%%%%%%%%%%%%%%%%%%%%%
%%%%%%%%%%%%%%%%%%%%%%%%%%%%%%%%%%%%%%%%%%%%%%%%%%%%%%%%%%%%%%%%%%%%%%%%

\section{Introduction}

The theory of general relativity (GTR), formulated by Albert
Einstein provides explanations for a broad range of gravitational
phenomena across various scales in the universe. Recent observations
confirming gravitational waves align with Einstein's predictions,
yet unresolved issues prompt exploration beyond GTR. One intriguing
avenue for addressing these challenges involves modifying GTR. A
notable example is $f(\mathfrak{R})$ gravity, a straightforward
modification with a substantial body of literature detailing its
realistic aspects. Additionally, the Lovelock theory of gravity
extends GTR to higher dimensions, named after mathematician David
Lovelock. Lovelock gravity retains GTR in four dimensions but offers
a more comprehensive description in higher dimensions. A distinctive
feature of Lovelock gravity is its prediction of black holes with
properties differing from those anticipated by GTR. Consequently,
Lovelock gravity has significance for investigating black holes and
other astrophysical phenomena. The Ricci scalar serves as the first
Lovelock scalar, while the second Lovelock scalar is \cite{3}
\begin{equation}\nonumber
\mathcal{G}=\mathfrak{R}_{\mu\nu\gamma\delta}\mathfrak{R}^{\mu\nu\gamma\delta}
+\mathfrak{R}^{2}-4\mathfrak{R}_{\mu\nu}\mathfrak{R}^{\mu\nu}.
\end{equation}
Nojiri and Odintsov \cite{4} formulated $f(\mathcal{G})$ gravity, a
theory devoid of instability issues \cite{5} while remaining
consistent with both solar system constraints \cite{6} and
cosmological structures \cite{7}. The introduction of
curvature-matter coupling was initially suggested in \cite{8}.
Subsequently, Harko et al \cite{9} established a similar coupling in
$f(\mathfrak{R})$ theory, referred to as $f(\mathfrak{R},T)$ theory.
Expanding on this, Sharif and Ikram \cite{10} extended the
$f(\mathcal{G})$ theory by incorporating the trace of the
stress-energy tensor in the functional action, resulting in
$f(\mathcal{G},T)$ gravity. These coupling mechanisms provide
explanations for phenomena such as the rotation curves of galaxies
and various evolutionary phases of the cosmos. Importantly, these
modified proposals lack conservation, indicating the presence of an
additional force and consequently altering the paths of particles.
Such modified theories prove valuable in elucidating the mysteries
of the dark universe.

Wormholes ($\mathrm{WH}$s) are theoretical constructs involving
linking two distinct points in spacetime, enabling the possibility
of faster-than-light travel or even time travel. If a $\mathrm{WH}$
connects different parts of separate universes, it is termed an
inter-universe $\mathrm{WH}$, whereas an intra-universe
$\mathrm{WH}$ links different regions within the same spacetime. The
$\mathrm{WH}$ takes the form of a tunnel-like structure connecting
two points in spacetime referred to as the mouths of the
$\mathrm{WH}$. The notion of a $\mathrm{WH}$ was initially
introduced by physicist Flamm in 1916. Later, Einstein and Rosen
explored the idea of a curved-space structure capable of uniting two
disparate cosmic regions through a bridge known as an Einstein-Rosen
bridge. Wheeler later determined that the Schwarzschild
$\mathrm{WH}$ is non-traversable. Moreover, the $\mathrm{WH}$ throat
rapidly expands from zero to a specific size and then contracts back
to zero, impeding access. It has been theorized that $\mathrm{WH}$s
would instantly collapse after formation.

The viability of a traversable $\mathrm{WH}$ geometry faces a
significant challenge arising from the need for an enormous amount
of exotic matter, which contradicts energy conditions. Morris and
Thorne \cite{15} were the pioneers in proposing a solution for a
traversable $\mathrm{WH}$, suggesting that exotic matter could be
confined at the throat of the $\mathrm{WH}$ through matching
constraints. Vacaru et al \cite{16} explored $\mathrm{WH}$ geometry
with anisotropic matter configurations, while Dzhunushaliev et al
\cite{17} investigated the impact of electromagnetic fields on the
stability of $\mathrm{WH}$ structures. In the realm of traversable
$\mathrm{WH}$ geometry, the study of wormhole shape functions (WSFs)
and their properties has emerged as an intriguing subject. Numerous
researchers \cite{18} have recently characterized the structure of
$\mathrm{WH}$s through various shape functions. Stability is a
critical factor governing how these hypothetical structures respond
to perturbations and enhance their physical properties. A
singular-free configuration leads to a stable $\mathrm{WH}$
structure, preventing collapse, while unstable $\mathrm{WH}$s can
exist due to very slow decay. Various methods have been employed to
analyze $\mathrm{WH}$ geometry, including constraints on fluid
parameters, specific forms of the equation of state, and the
solution of metric potentials. The embedding class-I method has been
introduced to establish a relation between temporal and radial
coordinates, facilitating the examination of celestial objects. This
method allows the embedding of an n-dimensional manifold into an
(n+m)-dimensional manifold. Recently, the study of spherical objects
with different matter distributions, guided by the Karmarkar
condition, has been explored \cite{19}-\cite{25}. Fayyaz and Shamir
\cite{26} conducted an examination of viable and stable
$\mathrm{WH}$ geometry using the Karmarkar condition.

Visser \cite{26a} developed traversable wormholes without the use of
symmetry. He aimed to generate the necessary exotic stress-energy by
examining the Casimir energy associated with oscillations of a
classical string. Poisson and Visser \cite{26b} used the linearized
equation of state to examine the viable and stable WH structures.
Maldacena and Maoz \cite{26c} developed Euclidean supergravity
solutions featuring multiple boundaries and observed the puzzling
nature of these configurations. Sushkov \cite{26d} used phantom
energy to examine the physical characteristics of static spherically
symmetric $\mathrm{WH}$ solutions and demonstrated that phantom
energy supports the stable WH structures. Lobo \cite{26e} developed
a thin shell around the interior $\mathrm{WH}$ solution through the
phantom equation of state. Maldacena and Susskind \cite{26f} found
that any pair of entangled black holes can be linked by the
Einstein-Rosen bridge. They examined that this connection extends
beyond black holes as in most cases the bridge may not have a smooth
geometric interpretation. Halilsoy et al \cite{26g} considered the
regular Hayward black hole to examine the thin shell wormholes. Ono
et al \cite{26h} explored the deflection angle of light for an
observer and source at finite distances from a rotating
$\mathrm{WH}$ in the weak field approximation. Ovgun et al
\cite{26i} discovered a new traversable $\mathrm{WH}$ solution in
the framework of a bumblebee gravity model, revealing that bumblebee
gravity effects lead to a non-trivial global topology of the
$\mathrm{WH}$ spacetime. Penington et al \cite{26j} investigated the
Page transition by summing replica geometries with different
topologies.

Visser \cite{27a} introduced a novel category of traversable
$\mathrm{WH}$s by merging two Schwarzschild spacetimes together and
analyzed that these constructed $\mathrm{WH}$s prevent the formation
of event horizons. Halilsoy and colleagues \cite{27b} investigated
the stability of thin-shell $\mathrm{WH}$s through linear
perturbations. Ovgun \cite{27c} used Darmois-Israel junction
conditions to analyze the exotic cosmic structures. Richarte et al
\cite{27d} developed traversable thin-shell $\mathrm{WH}$s by using
the cut and paste technique. Ovgun \cite{27e} used the Gauss-Bonnet
theorem to analyze weak gravitational lensing in
rotating/non-rotating Damour-Solodukhin $\mathrm{WH}$s. Jusufi and
Ovgun \cite{27f} determined the deflection angle of rotating
$\mathrm{WH}$s using the Gauss-Bonnet theorem. Ovgun \cite{27h}
studied the viable $\mathrm{WH}$ structures through dark matter
medium. Kumaran and Ovgun \cite{27i} utilized the Gauss-Bonnet
theorem to derive the weak deflection angle for traversable
$\mathrm{WH}$s. Ovgun \cite{27j} studied the behavior of evolving
topologically deformed $\mathrm{WH}$s supported in dark matter halos
and examined their characteristics during the inflation era. Javed
et al \cite{27k} explored a $\mathrm{WH}$-like static aether
solution and computed the deflection angle in various mediums.

The study of $\mathrm{WH}$s presents intriguing outcomes in modified
theories. Lobo and Oliveira \cite{27} delved into $\mathrm{WH}$
structures by examining equations of state in the context of
$f(\mathfrak{R})$ theory. Azizi \cite{28} conducted an analysis of
the geometry of wormholes, employing various shape functions in the
realm of $f(\mathfrak{R},\mathcal{T})$ theory. Sharif and Fatima
\cite{29} explored $\mathrm{WH}$ structures in $f(\mathcal{G})$
theory, considering both constant and variable redshift functions.
Elizalde and Khurshudyan \cite{29a} assumed a barotropic equation of
state to assess the viability of traversable $\mathrm{WH}$s in
$f(\mathfrak{R},\mathcal{T})$ theory. Sharif and Hussain \cite{30}
utilized the Noether symmetry approach to investigate static
spherically symmetric $\mathrm{WH}$ solutions in the framework of
$f(\mathcal{G},\mathcal{T})$ gravity. Shamir and Fayyaz \cite{33}
discovered that a minimal amount of exotic matter could give rise to
$\mathrm{WH}$ geometry in $f(\mathfrak{R})$ theory. We have explored
$\mathrm{WH}$ solutions under various considerations in
$f(\mathfrak{R},\mathcal{T}^2)$ theory \cite{34}. Malik et al
\cite{36} applied the Karmarkar technique to analyze static
spherical solutions in the context of $f(\mathfrak{R})$ theory.

In this study, we employ the embedding class-I approach to
investigate viable traversable solutions in the framework of
$f(\mathcal{G},T)$ theory. The analysis focuses on examining the
behavior of the null energy condition and shape function in this
context. The structure of the paper is as follows. Section
\textbf{2} explores the derivation of the WSF using the Karmarkar
condition. In Section \textbf{3}, we formulate the field equations
for a static spherical spacetime in $f(\mathcal{G},T)$ theory and
examine the null energy condition corresponding to various models of
the theory. Section \textbf{4} assesses the stability of the
$\mathrm{WH}$ solutions through causality conditions, the Herrera
cracking approach and the adiabatic index. The last section provides
a summary of our findings.

\section{Wormhole Geometry}

\begin{figure}
\epsfig{file=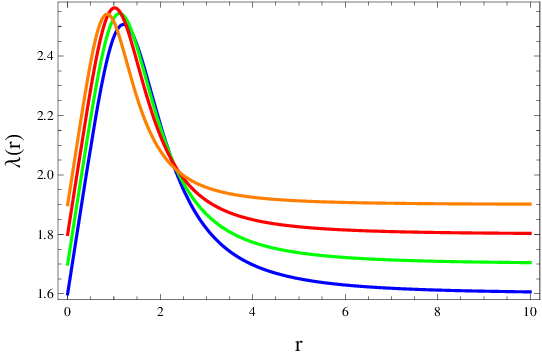,width=.5\linewidth}\epsfig{file=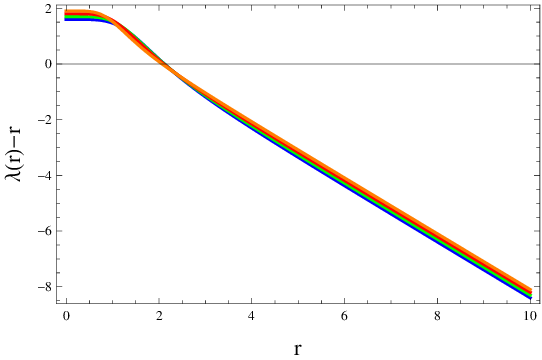,width=.5\linewidth}
\epsfig{file=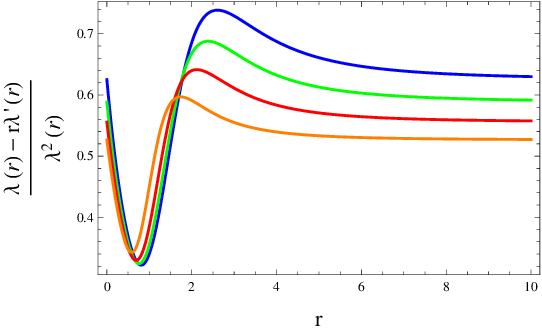,width=.5\linewidth}\epsfig{file=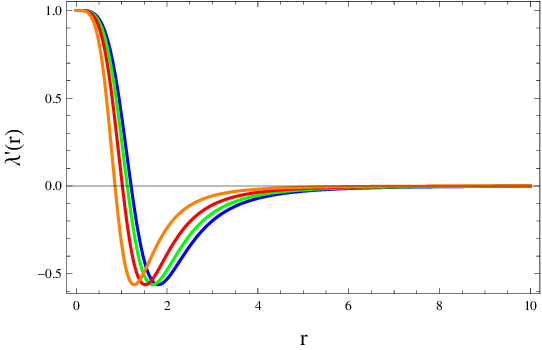,width=.5\linewidth}
\epsfig{file=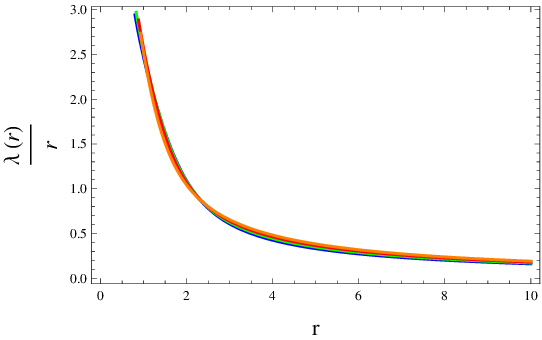,width=.5\linewidth}\caption{\label{F1}Behavior
of WSF for $a=1.9$ (blue), $a=1.8$ (green), $a=1.7$ (red) and
$a=1.6$ (orange).}
\end{figure}

We consider static spherical metric as
\begin{equation}\label{1}
ds^{2}=-dt^{2}e^{\xi(r)}+dr^{2}e^{\eta(r)}+d\theta^{2}r^{2}
+d\phi^{2}\sin^{2}\theta.
\end{equation}
The Karmarkar condition is defined as
\begin{eqnarray}\label{2}
\mathfrak{R}_{1414}&=&\frac{\mathfrak{R}_{1212}\mathfrak{R}_{3434}
+\mathfrak{R}_{1224}\mathfrak{R}_{1334}}{\mathfrak{R}_{2323}},\quad
\mathfrak{R}_{2323}\neq0.
\end{eqnarray}
where
\begin{eqnarray}\nonumber
\mathfrak{R}_{1212}&=&\frac{e^{\xi}(2\xi''+\xi'^{2}-\xi'\eta')}{4},
\quad
\mathfrak{R}_{3434}=\frac{r^{2}\sin^{2}\theta(e^{\eta}-1)}{e^{\eta}},
\quad \mathfrak{R}_{1414}=\frac{r\sin^{2}\theta\xi'e^{\xi-\eta}}{2},
\\\nonumber
\mathfrak{R}_{2323}&=&\frac{r\eta'}{2}, \quad
\mathfrak{R}_{1334}=\mathfrak{R}_{1224}\sin^{2}\theta.
\end{eqnarray}
Solving this constraint, we obtain
\begin{equation}\nonumber
\frac{\xi'\eta'}{1-e^{\eta}}=\xi'\eta'-2\xi''-\xi'^{2},
\end{equation}
where $e^{\eta}\neq1$. The corresponding solution with integration
constant $\alpha$ is
\begin{equation}\label{3}
e^{\eta}=1+\alpha e^{\xi}\xi'^{2},
\end{equation}
To develop the WSF, we consider the Morris-Thorne spacetime as
\begin{equation}\label{4}
ds^{2}=-dt^{2}e^{\xi(r)}+dr^{2}\frac{1}{1-\frac{\lambda(r)}{r}}
+d\theta^{2}r^{2}+d\phi^{2}r^{2}\sin\theta.
\end{equation}
Here, $\lambda$ is the shape function and the redshift function is
defined as \cite{37}
\begin{equation}\label{5}
\xi(r)=\frac{-2\beta}{r},
\end{equation}
where $\beta$ is an arbitrary constant. Using Eqs.(\ref{1}) and
(\ref{4}) gives
\begin{equation}\label{6}
\eta(r)=\ln\left[\frac{r}{r-\lambda(r)}\right].
\end{equation}
Equations (\ref{3}) and (\ref{5}) gives
\begin{equation}\label{7}
\lambda(r)=r-\frac{r^{5}}{r^{4}+4\beta^{2}\alpha
e^{\frac{-2\beta}{r}}}.
\end{equation}
For a traversable $\mathrm{WH}$ solution, Morris and Thorne
\cite{38} stated that shape function must satisfy the following
conditions
\begin{enumerate}
\item
$\lambda(r)<r$,
\item
$\lambda(r)-r=0$ at $r=r_{0}$,
\item
$\frac{\lambda(r)-r\lambda'(r)}{\lambda^{2}(r)}>0$ at $r=r_{0}$,
\item
$\lambda'(r)<1$,
\item
$\frac{\lambda(r)}{r}\rightarrow0$ when $r\rightarrow\infty$,
\end{enumerate}
where $r_{0}$ is the radius of $\mathrm{WH}$ throat. Equation
(\ref{7}) has a trivial solution at $\mathrm{WH}$ throat, i.e.,
$\lambda(r_{0})-r_{0}=0$. Therefore, we redefine Eq.(\ref{7}) for
non-trivial solution as
\begin{eqnarray}\label{8}
\lambda(r)=r-\frac{r^{5}}{r^{4}+4\beta^{2}\alpha
e^{\frac{-2\eta}{r}}}+a, \quad 0<a<r_{0}.
\end{eqnarray}
Using the condition (2), we have
\begin{equation}\label{9}
\beta=\frac{r_{0}^{4}(r_{0}-a)}{4\alpha^{2}e^{\frac{-2\alpha^{2}}{r_{0}}}}.
\end{equation}
Substituting the value of $\beta$ in Eq.(\ref{8}), one can obtain
the expression of shape function as
\begin{eqnarray}\label{10}
\lambda(r)=r-\frac{r^{5}}{r^{4}+r_{0}^{4}(r_{0}-a)}+a, \quad
0<a<r_{0}.
\end{eqnarray}
Conditions (3) and (4) fulfill for the given range of $a$. Applying
condition (5) on Eq.\eqref{10}, we have
\begin{equation}\label{10a}
\lim_{r\rightarrow\infty}\frac{\lambda(r)}{r}=0.
\end{equation}
We assume $r_{0}=2$ and $\eta=-1$ for our convenience in all the
graphs. The graphical representation of the WSF is given in Figure
\ref{F1} which manifests that the WSF satisfies all the required
conditions.

\section{Modified Gravitational Theory}

The corresponding integral action is defined as \cite{10}
\begin{equation}\label{12}
\mathcal{S}=\frac{1}{2\kappa}\int\left[\mathfrak{R}+f
(\mathcal{G},T)\right]{\sqrt{-g}}d^4x+\int
\mathcal{L}_{m}\sqrt{-g}d^4x,
\end{equation}
where $\mathcal{L}_{m}$ represents matter-Lagrangian density and $g$
is the determinant of the metric tensor. The corresponding field
equations are
\begin{eqnarray}\nonumber
G_{\mu\nu}&=& 8\pi T_{\mu\nu}-(\Theta_{\mu\nu}+T_{\mu\nu})
f_{T}+\frac{1}{2}g_{\mu\nu}f+\big(4\mathfrak{R}_{\gamma\nu}
\mathfrak{R}^{\gamma}_{\mu}+4\mathfrak{R}^{\gamma\delta}
\mathfrak{R}_{\mu\gamma\nu\delta}-2\mathfrak{R}\mathfrak{R}_{\mu\nu}-2
\mathfrak{R}_{\nu\gamma\delta\zeta}\mathfrak{R}_{\mu}^{\gamma\delta\zeta}\big)
f_{\mathcal{G}}+\big(4\mathfrak{R}_{\mu\nu}\nabla_{2}
\\\label{13}
&+&4g_{\mu\nu}\mathfrak{R}^{\gamma\delta}\nabla_{\gamma}
\nabla_{\delta}+2\mathfrak{R}\nabla_{\mu}\nabla_{\nu}-2
g_{\mu\nu}\mathfrak{R}\nabla^{2}
-4\mathfrak{R}^{\gamma}_{\mu}\nabla_{\nu}\nabla_{\gamma}
-4\mathfrak{R}^{\gamma}_{\nu}\nabla_{\mu}\nabla_{\gamma}
-4\mathfrak{R}_{\mu\gamma\nu\delta}
\nabla^{\gamma}\nabla^{\delta}\big)f_{\mathcal{G}}.
\end{eqnarray}
Here, $f\equiv f(\mathcal{G},T)$, $f_{\mathcal{G}}= \frac{\partial
f} {\partial \mathcal{G}}$, $f_{\mathfrak{R}}= \frac{\partial f}
{\partial \mathfrak{R}}$ and the field equations of this theory
reduce to $f(\mathcal{G})$ gravity for
$f(\mathcal{G},T)=f(\mathcal{G})$. The expression of
$\Theta_{\mu\nu}$ is given by
\begin{equation}\label{14}
\Theta
_{\mu\nu}=-2T_{\mu\nu}+g_{\mu\nu}\mathcal{L}_{m}-2g^{\gamma\delta}
\frac{\partial^{2}\mathcal{L}_{m}}{\partial g^{\alpha\beta}\partial
g^{\gamma\delta}}.
\end{equation}

We assume matter distribution as
\begin{equation}\label{15}
T_{\mu\nu}=
\mathcal{U}_{\mu}\mathcal{U}_{\nu}(\rho+P_{t})-P_{t}g_{\mu\nu}
+\mathcal{V}_{\mu}\mathcal{V}_{\nu}(P_{r}-P_{t}),
\end{equation}
where $\mathcal{U}_{\mu}$ and $\mathcal{V}_{\mu}$ defines the
four-velocity and four-vector of the fluid, respectively. Here, we
use specific model of this theory as \cite{39}
\begin{equation}\label{16}
f(\mathcal{G},T)=f_{1}(\mathcal{G})+f_{2}(T).
\end{equation}
This model gives a suitable extension of $f(\mathcal{G})$ theory.
The viable models of $f(\mathcal{G},T)$ can be developed by
considering different forms of $f_{1}(G)$ with $f_{2}(T)=\gamma T$,
where $\gamma$ is an arbitrary constant. The resulting field
equations are
\begin{eqnarray}\nonumber
\rho&=&\frac{e^{-2\eta}}{8r^{4}(1+\gamma)(1+2\gamma)}\bigg[8e^{\eta}
(e^{\eta}-1)r^2(1+2\gamma)-4r^{4}e^{2\eta}r^{4}(1+\gamma)f_{1}
+f_{\mathcal{G}}\bigg[-16(e^{\eta}-1)^{2}\gamma+r^{2}\bigg\{r^{2}
(1+2\gamma)\xi'^{4}
\\\nonumber
&-&2r^{2}(1+2\gamma)\xi'^{3}\eta'-4\xi'\eta'\big\{2(e^{\eta}-3)(1+\gamma)
+r^{2}(1+2\gamma)\xi''+\xi'^{2}\left(8(\gamma+e^{\eta}(1+\gamma))\right)
+r^{2}(1+2\gamma)(\eta'^{2}+4\xi'')\big\}
\\\nonumber
&+&4\big\{-2\gamma
\eta'^{2}+\xi''(4(e^{\eta}-1)(1+\gamma))\big\}+r^{2}(1+2\gamma)\xi''\bigg\}
\bigg]+2r\bigg\{4f_{\mathcal{G}}'
\big\{-8(2+5\gamma)+r(\eta'(10+27\gamma-2r\gamma \eta'))
\\\nonumber
&-&r(8+18\gamma+r(2+3\gamma)\eta')\xi''\big\}-8r(2+5\gamma)(1-2r\eta'
+r^{2}\xi'')f_{\mathcal{G}}''
+r^{2}\xi'^{2}\big\{re^{\eta}\gamma-2(8+18\gamma+r(2+3\gamma)\eta')
f_{\mathcal{G}}'
\\\nonumber
&-&4r(2+5\gamma)f_{\mathcal{G}}''\big\}+2e^{\eta}\big\{16(2+5\gamma)
f_{\mathcal{G}}'
+2r\eta'(r+2r\gamma+(2+3\gamma)f_{\mathcal{G}}')+r^{3}\gamma
\xi''+4r(2+5\gamma)f_{\mathcal{G}}''\big\}+r\xi'\big\{2(-32
\\\label{17}
&-&74\gamma+r\eta'(2\gamma+r(2+3\gamma)\eta'))\big\}f_{\mathcal{G}}'
-e^{\eta}\gamma(r(-4+r\eta')+4f_{\mathcal{G}}')
+4r\big\{-2(4+9\gamma)+r(2+5\gamma)\eta'f_{\mathcal{G}}''\big\}
\bigg\}\bigg],
\\\nonumber
P_{r}&=&\frac{e^{-2\eta}}{8r^{4}(1+\gamma)(1+2\gamma)}\bigg[-4e^{2\eta}
r^{4}(1+\gamma)f_{1}
+f_{\mathcal{G}}\bigg[-16(e^{\eta}-1)^{2}\gamma+r^{2}
\bigg\{r^{2}(1+2\gamma)\xi'^{4}-2r^{2}(1+2\gamma)\xi'^{3}\eta'-4\xi'\eta'
\\\nonumber
&\times&(2(e^{\eta}-3)(1+\gamma)
+r^{2}(1+2\gamma)\xi'')+\xi'^{2}\left\{8(e^{\eta}-1)(1+\gamma)
+r^{2}(1+2\gamma)(\eta'^{2}+4\xi'')\right\}
+4\big\{2(1+\gamma)\eta'^{2}
\\\nonumber
&+&\xi''(4(e^{\eta}-1)(1+\gamma)+r^{2}(1+2\gamma)\xi'')
\big\}\bigg\}\bigg]
+2r(4e^{\eta}(e^{\eta}-1))r(1+2\gamma)+4f_{\mathcal{G}}'
\bigg[-8\gamma+r\bigg\{-\eta'(4+\gamma+2r\gamma
\eta')
\\\nonumber
&+&r\gamma(-2+r \eta'\xi'')+r^{2}\gamma
\xi'^{2}(2(-2+r\eta')f_{\mathcal{G}}'+r(e^{\eta}-4f_{\mathcal{G}}''))
-8r\gamma(1-2r\eta'+r^{2}\xi'')f_{\mathcal{G}}''
+r\xi'\big\{-2\big(12+34\gamma
\\\nonumber
&+&r\eta'(8+14\gamma+4\gamma
\eta')\big)f_{\mathcal{G}}'+e^{\eta}(-r(4+4\gamma+r\gamma
\eta')+4(2+3\gamma)f_{\mathcal{G}}')
+4r\gamma(-2+r\eta')f_{\mathcal{G}}''\big\}+2e^{\eta}
\big(2(8\gamma+r(4+7\gamma)
\\\label{18}
&\times&\eta')f_{\mathcal{G}}'+r\gamma(r^2\xi''+4f_{\mathcal{G}}'')
\big)\bigg\}\bigg]\bigg],
\\\nonumber
p_{t}&=&\frac{e^{-2\eta}}{4r^{4}(1+\gamma)(1+2\gamma)}
\bigg[-2e^{2\eta}r^{4}(1+\gamma)f_{1}
+2f_{\mathcal{G}}\bigg\{4(e^{\eta}-1)^{2}(1+\gamma)+r^{2}
\big\{(-1+2e^{\eta}(1+\gamma))a'^{2}-2(-3+e^{\eta})
\\\nonumber
&\times&(1+\gamma)\xi'\eta'+\eta'^{2}+4(-1+e^{\eta})(1+\gamma)
\xi''\big\}\bigg\}
-r\bigg\{4f_{\mathcal{G}}'\big\{8\gamma+r(\eta'(-7\gamma-2r(1+\gamma)\eta')
+r(2+6\gamma+r(2+3\gamma)
\\\nonumber
&\times&\eta')\xi'')\big\}+8r\gamma(1-2r\eta'+r^{2}\xi'')
f_{\mathcal{G}}''+2e^{\eta}\left\{-16\gamma
f_{\mathcal{G}}'-r\eta'(r+2r\gamma-2\gamma
f_{\mathcal{G}}')+r^{3}(1+\gamma)\xi''-4r\gamma
f_{\mathcal{G}}''\right\}+r^{2}\xi'^{2}
\\\nonumber
&\times&\big\{e^{\eta}r(1+\gamma)+2\left(2+6\gamma+r(2+3\gamma)
\eta'\right)f_{\mathcal{G}}'+4r\gamma
f_{\mathcal{G}}''\big\}+r\xi'-2\left(-10\gamma+r\eta'
(2+6\gamma+r(2+3\gamma)\eta')
\right)f_{\mathcal{G}}'+e^{\eta}
\\\label{19}
&\times&\big(-r(-2+r(1+\gamma)\eta')+4\gamma
f_{\mathcal{G}}'\big)+4r(2+6\gamma-r\gamma
\eta')f_{\mathcal{G}}''\bigg\}\bigg].
\end{eqnarray}

\subsection{Model l}

\begin{figure}
\epsfig{file=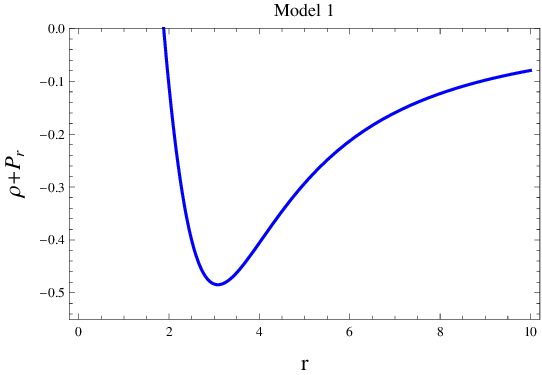,width=.5\linewidth}\epsfig{file=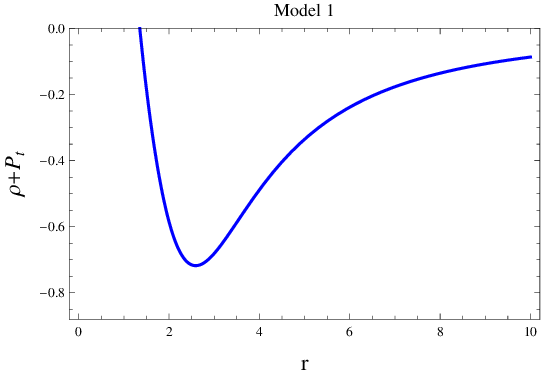,width=.5\linewidth}
\epsfig{file=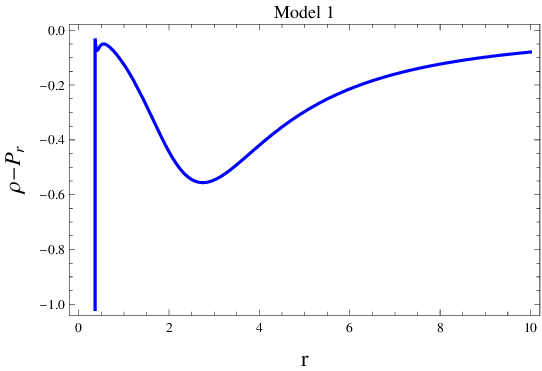,width=.5\linewidth}\epsfig{file=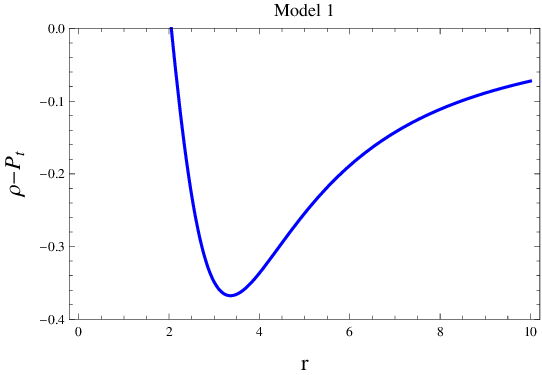,width=.5\linewidth}
\epsfig{file=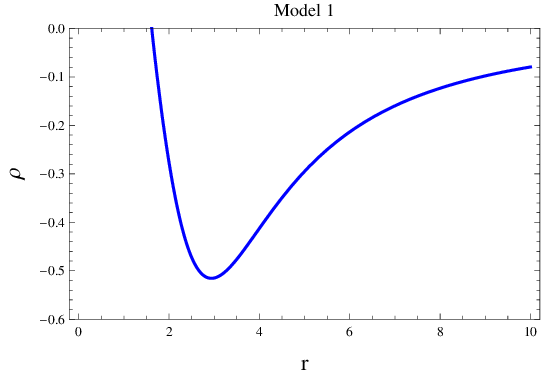,width=.5\linewidth}\epsfig{file=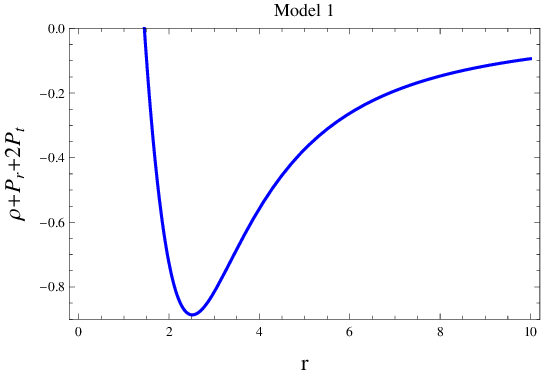,width=.5\linewidth}
\caption{\label{F2}Graphs of energy bounds for $a_{1}=0.0001$,
$b_{1}=0.00001$, $n_{1}=2$ and $\gamma=0.001$.}
\end{figure}

\begin{figure}
\epsfig{file=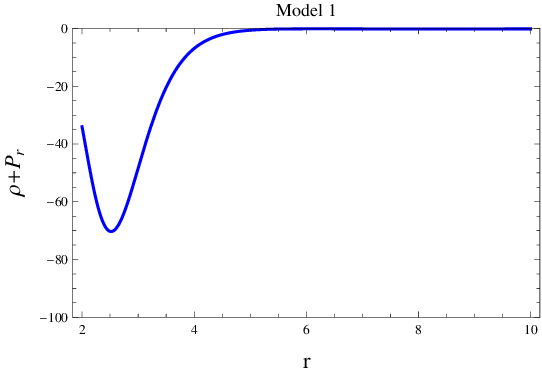,width=.5\linewidth}\epsfig{file=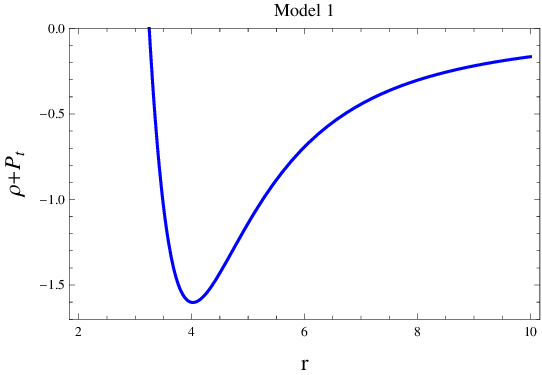,width=.5\linewidth}
\epsfig{file=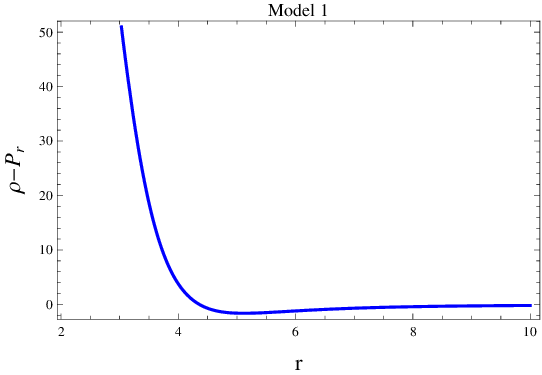,width=.5\linewidth}\epsfig{file=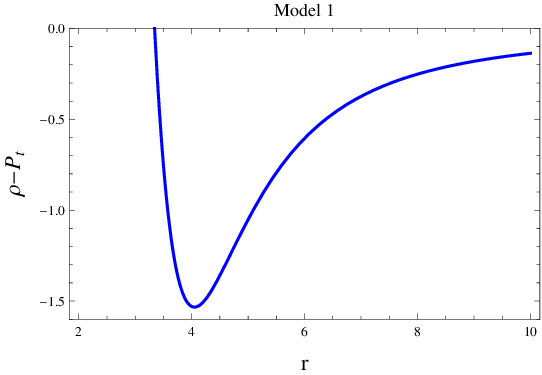,width=.5\linewidth}
\epsfig{file=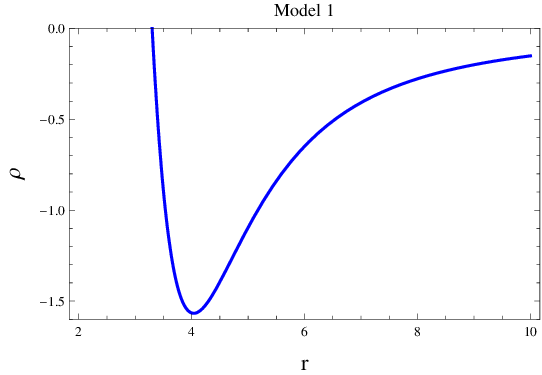,width=.5\linewidth}\epsfig{file=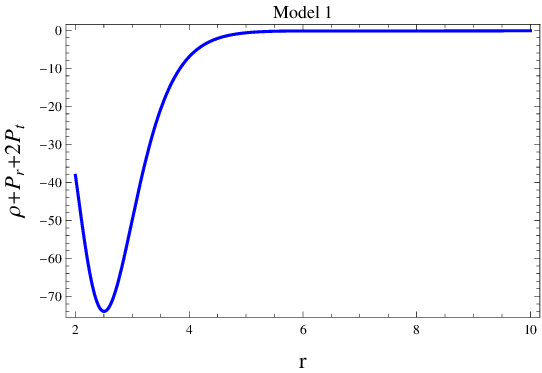,width=.5\linewidth}
\caption{\label{F3}Graphs of energy bounds for $a_{1}=3$, $b_{1}=7$,
$n_{1}=5$ and $\gamma=0.001$.}
\end{figure}

\begin{figure}
\epsfig{file=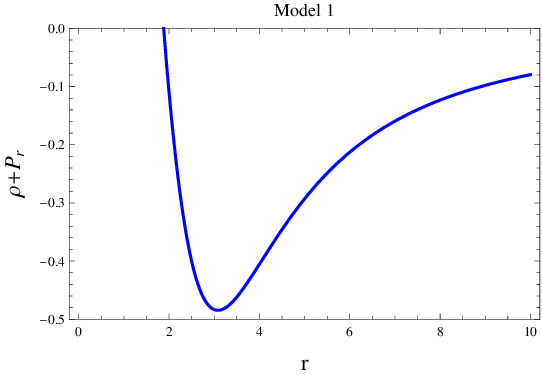,width=.5\linewidth}\epsfig{file=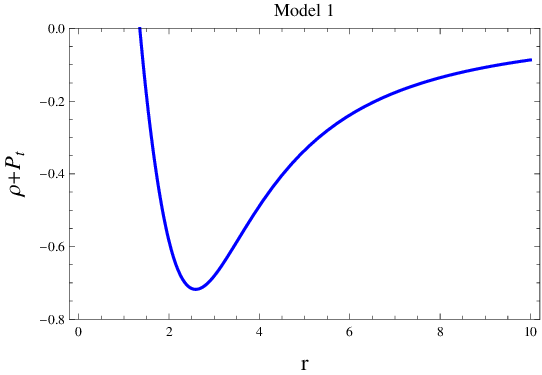,width=.5\linewidth}
\epsfig{file=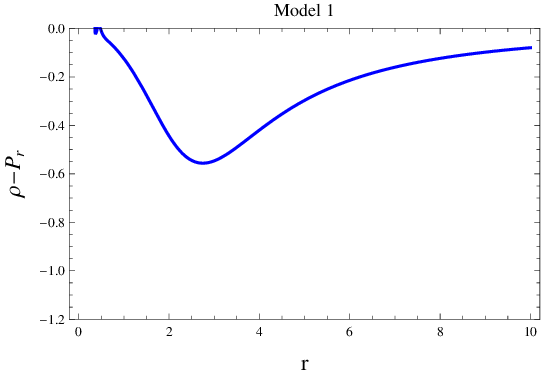,width=.5\linewidth}\epsfig{file=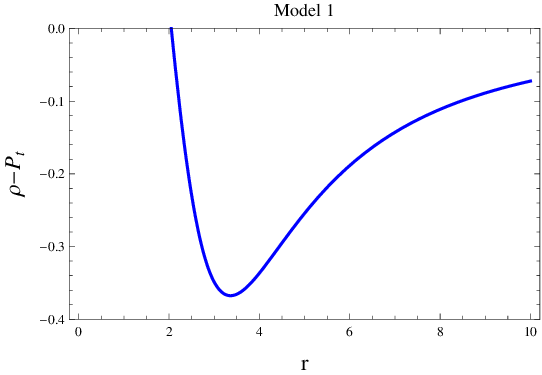,width=.5\linewidth}
\epsfig{file=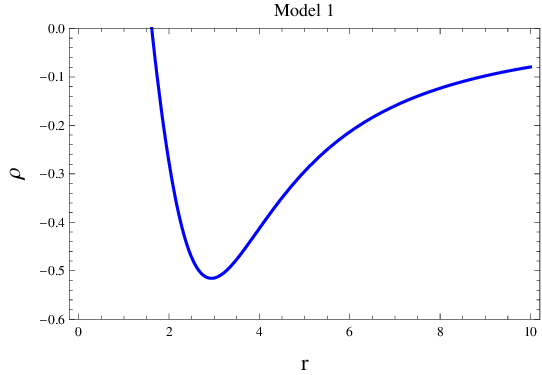,width=.5\linewidth}\epsfig{file=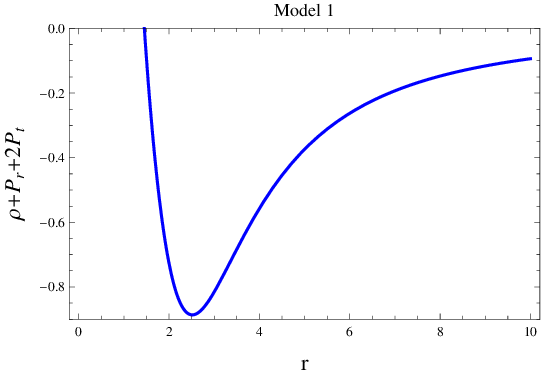,width=.5\linewidth}
\caption{\label{F4}Graphs of energy bounds for $a_{1}=-0.0001$,
$b_{1}=-0.00001$, $n_{1}=7$ and $\gamma=0.001$.}
\end{figure}

\begin{figure}
\epsfig{file=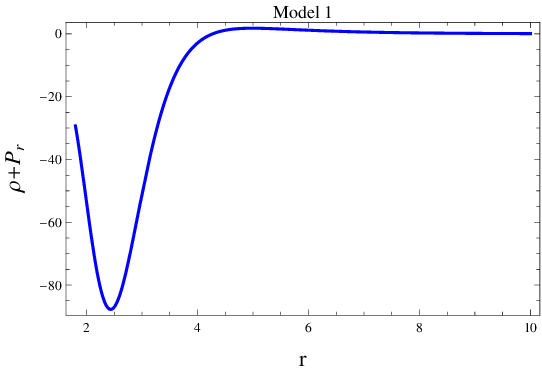,width=.5\linewidth}\epsfig{file=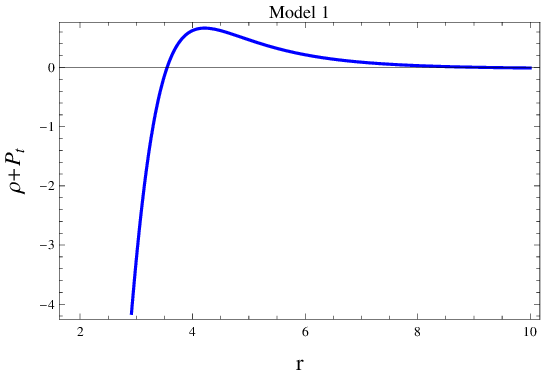,width=.5\linewidth}
\epsfig{file=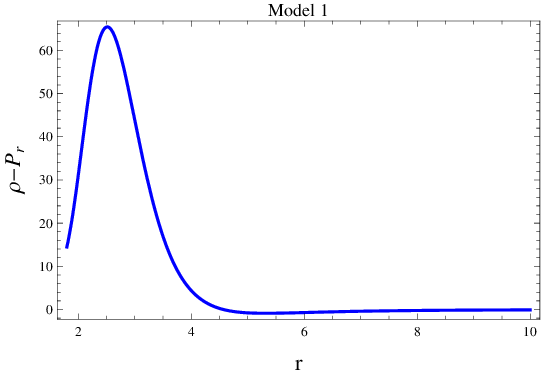,width=.5\linewidth}\epsfig{file=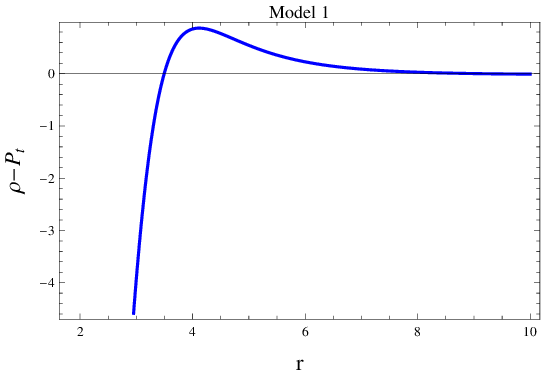,width=.5\linewidth}
\epsfig{file=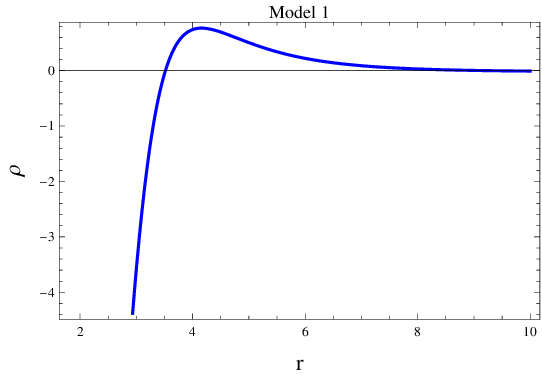,width=.5\linewidth}\epsfig{file=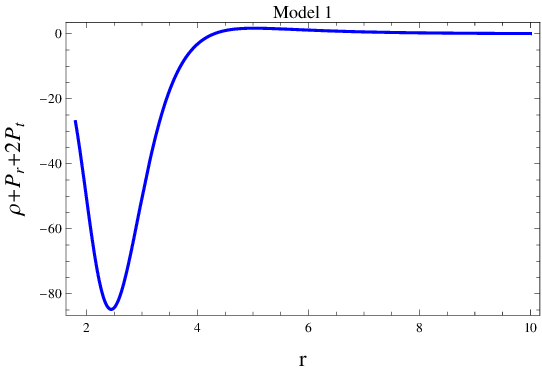,width=.5\linewidth}
\caption{\label{F5}Graphs of energy bounds for $a_{1}=-3$,
$b_{1}=-7$, $n_{1}=9$ and $\gamma=0.001$.}
\end{figure}

\begin{figure}
\epsfig{file=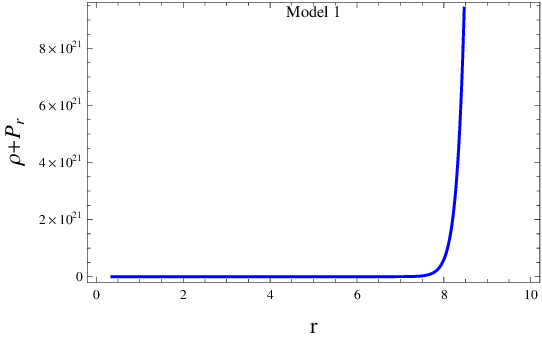,width=.5\linewidth}\epsfig{file=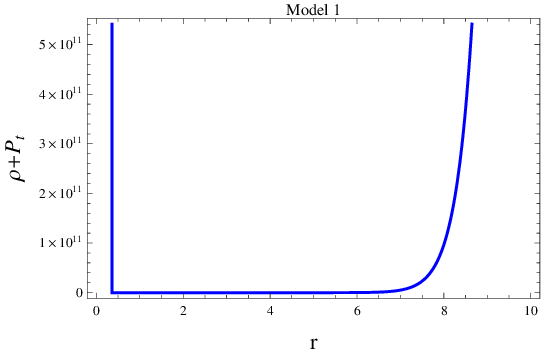,width=.5\linewidth}
\epsfig{file=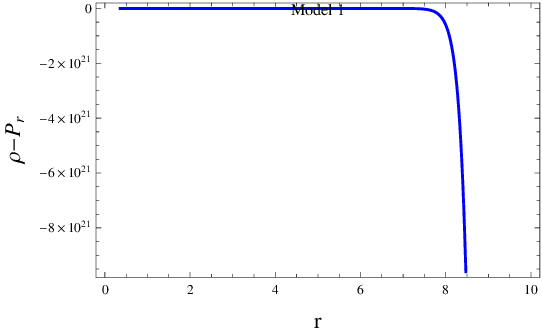,width=.5\linewidth}\epsfig{file=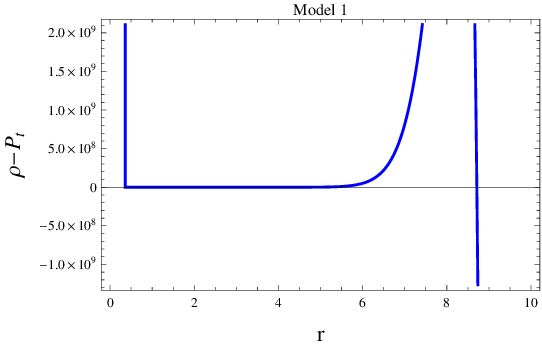,width=.5\linewidth}
\epsfig{file=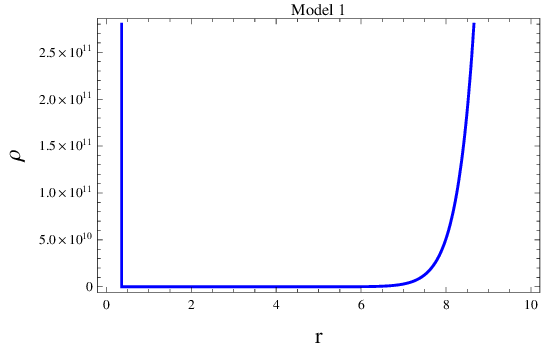,width=.5\linewidth}\epsfig{file=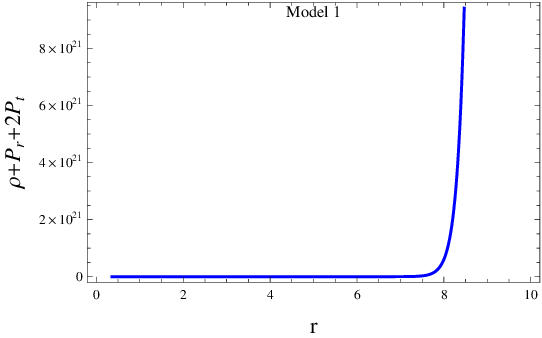,width=.5\linewidth}
\caption{\label{F6}Graphs of energy bounds for $a_{1}=0.0001$,
$b_{1}=0.00001$, $n_{1}=-2$ and $\gamma=0.001$.}
\end{figure}

\begin{figure}
\epsfig{file=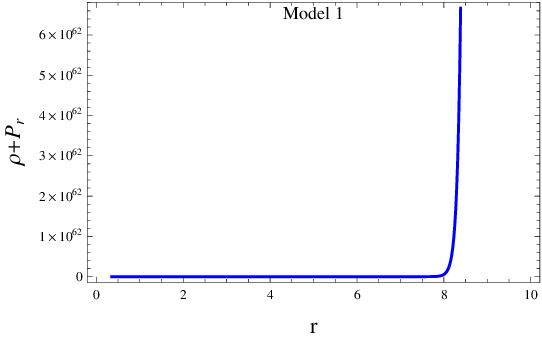,width=.5\linewidth}\epsfig{file=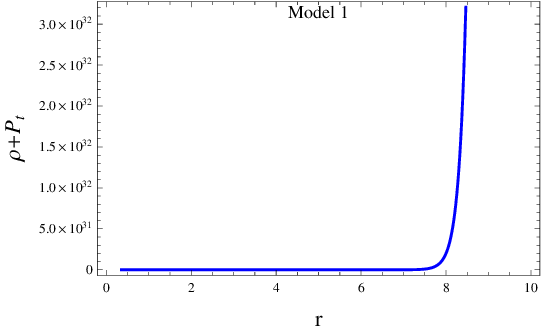,width=.5\linewidth}
\epsfig{file=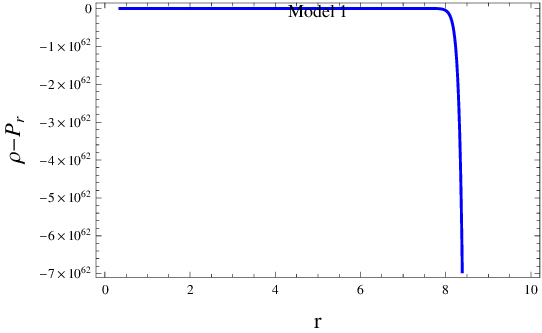,width=.5\linewidth}\epsfig{file=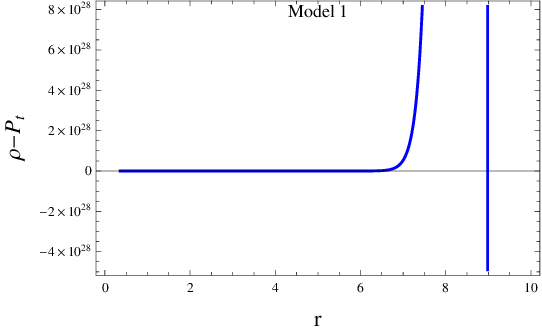,width=.5\linewidth}
\epsfig{file=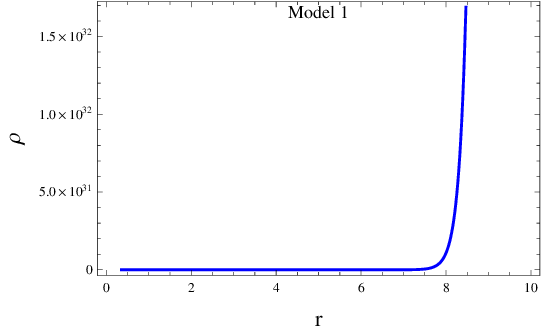,width=.5\linewidth}\epsfig{file=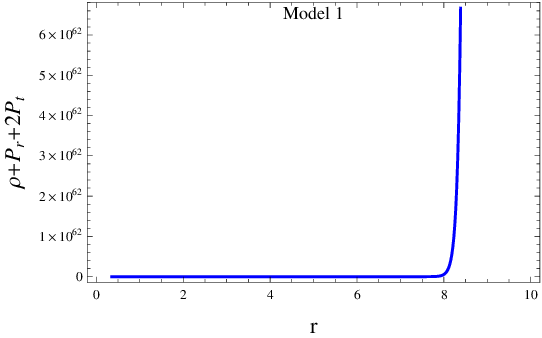,width=.5\linewidth}
\caption{\label{F7}Graphs of energy bounds for $a_{1}=3$, $b_{1}=7$,
$n_{1}=-5$ and $\gamma=0.001$.}
\end{figure}

\begin{figure}
\epsfig{file=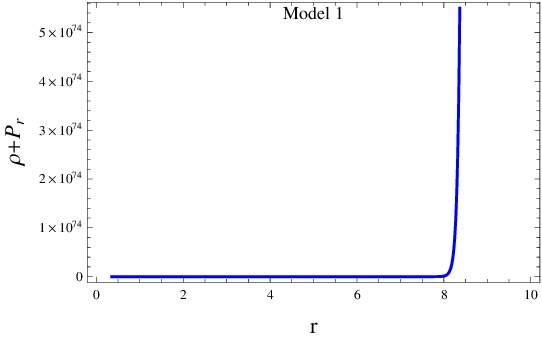,width=.5\linewidth}\epsfig{file=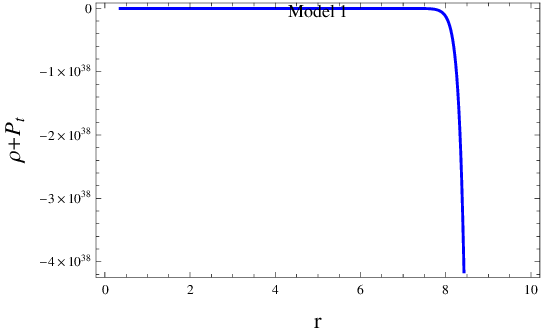,width=.5\linewidth}
\epsfig{file=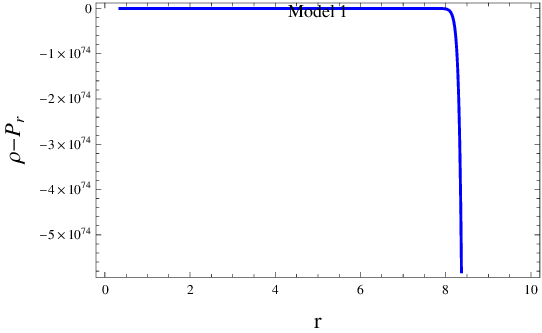,width=.5\linewidth}\epsfig{file=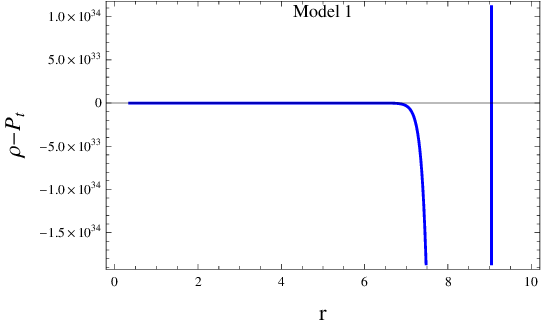,width=.5\linewidth}
\caption{\label{F8}Graphs of energy bounds for $a_{1}=-0.0001$,
$b_{1}=-0.00001$, $n_{1}=-7$ and $\gamma=0.001$.}
\end{figure}

\begin{figure}
\epsfig{file=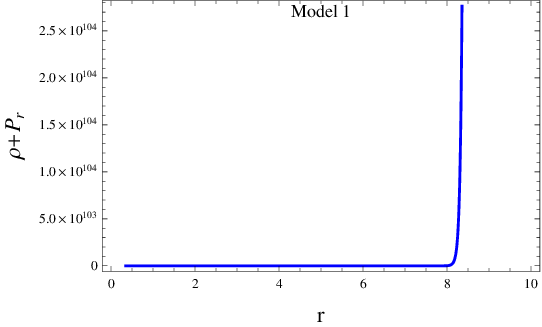,width=.5\linewidth}\epsfig{file=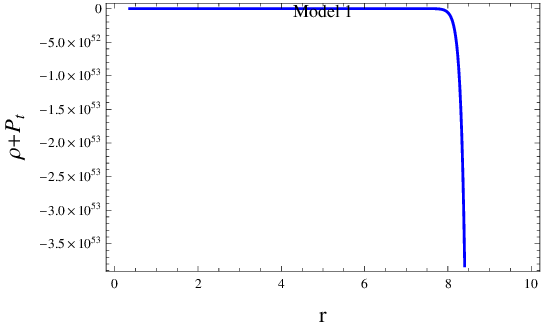,width=.5\linewidth}
\epsfig{file=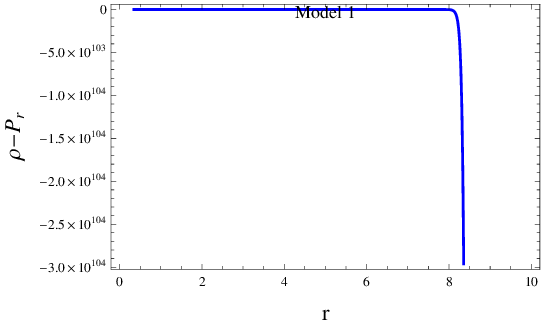,width=.5\linewidth}\epsfig{file=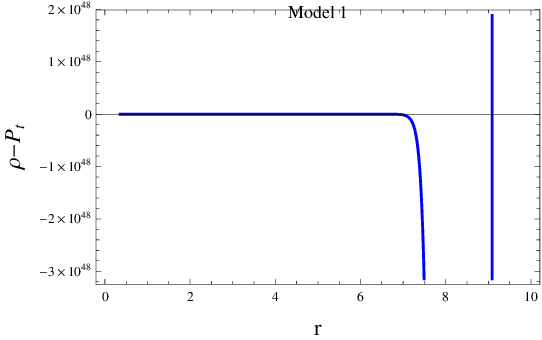,width=.5\linewidth}
\epsfig{file=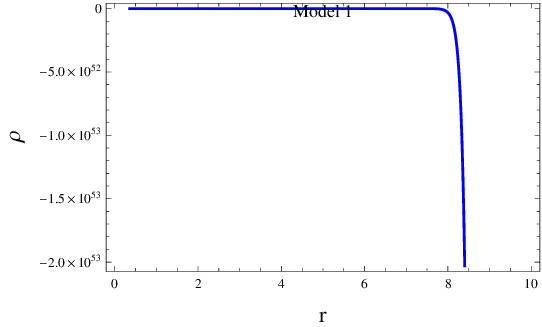,width=.5\linewidth}\epsfig{file=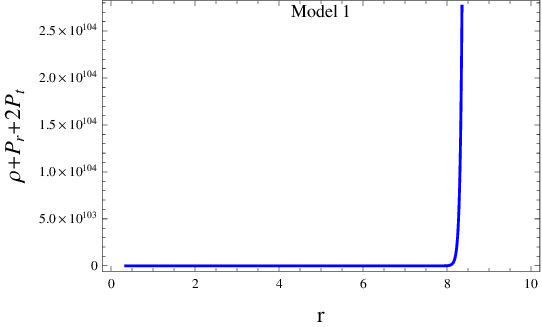,width=.5\linewidth}
\caption{\label{F9}Graphs of energy bounds for $a_{1}=-3$,
$b_{1}=-7$, $n_{1}=-9$ and $\gamma=0.001$.}
\end{figure}

We first consider the power-law model as \cite{40}
\begin{equation}\label{20}
f(\mathcal{G},T)=a_{1}\mathcal{G}^{n_{1}}+b_{1}\mathcal{G}\ln(\mathcal{G})+\gamma
T,
\end{equation}
where $a_{1}$, $b_{1}$ and $n_{1}$ are arbitrary constants. The
corresponding field equations are given in Appendix \textbf{A}.
Energy conditions play an important role in determining the presence
of cosmic structures. These conditions must be violated for a viable
$\mathrm{WH}$ structure. Modified theories of gravity ensure that
the viable traversable $\mathrm{WH}$ geometry exists if the energy
conditions are violated. The graphical behavior of energy conditions
for distinct values of model parameters is given in Figures
\ref{F2}-\ref{F9}. These graphs ensure the existence of a viable
traversable $\mathrm{WH}$ geometry for all values of $a_{1}$,
$b_{1}$ and $n_{1}$.

\subsection{Model 2}

\begin{figure}
\epsfig{file=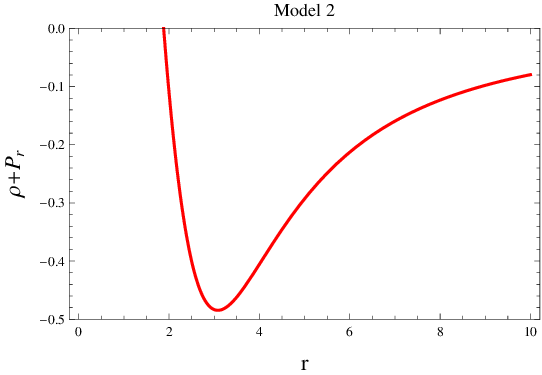,width=.5\linewidth}\epsfig{file=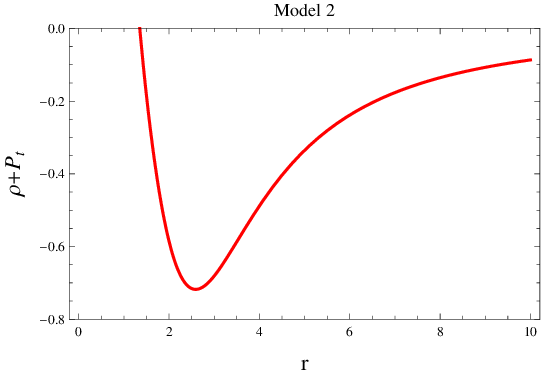,width=.5\linewidth}
\epsfig{file=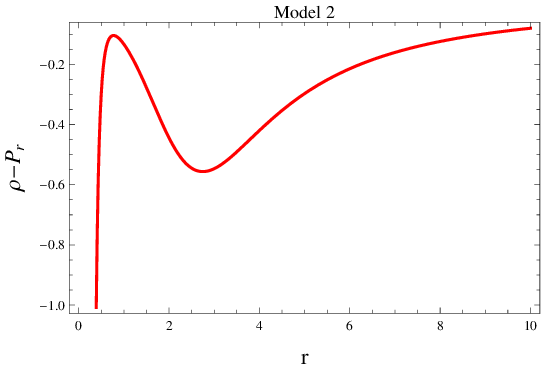,width=.5\linewidth}\epsfig{file=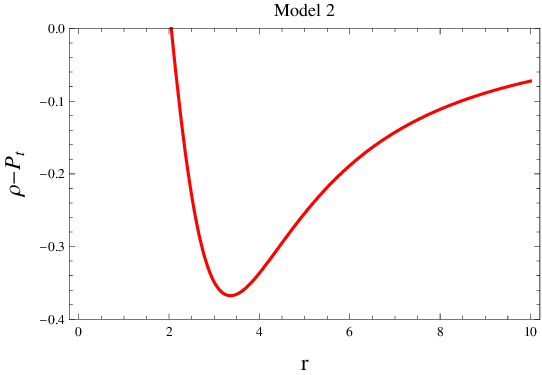,width=.5\linewidth}
\epsfig{file=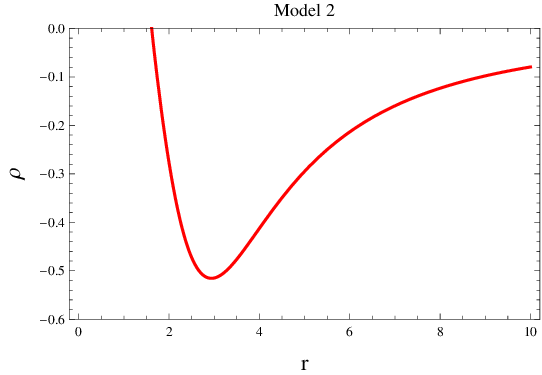,width=.5\linewidth}\epsfig{file=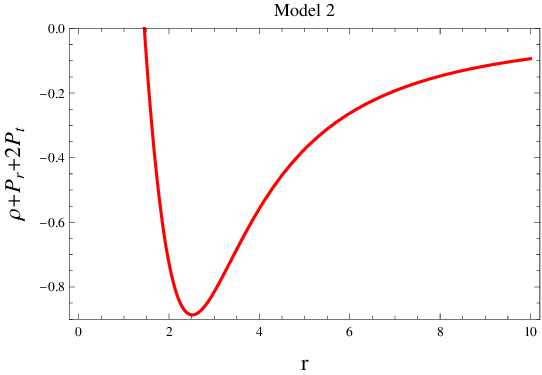,width=.5\linewidth}
\caption{\label{F10}Graphs of energy bounds for $a_{2}=0.002$,
$b_{2}=0.0003$, $n_{2}=1$, $m=0.0003$ and $\gamma=0.001$.}
\end{figure}

\begin{figure}
\epsfig{file=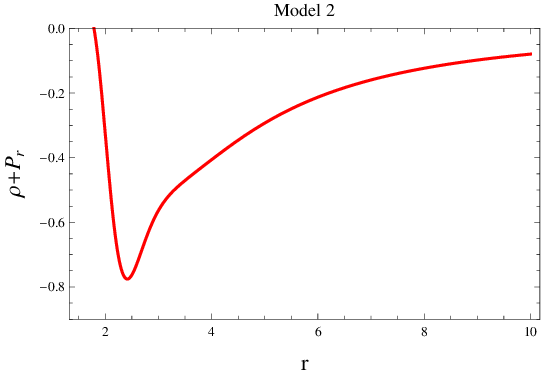,width=.5\linewidth}\epsfig{file=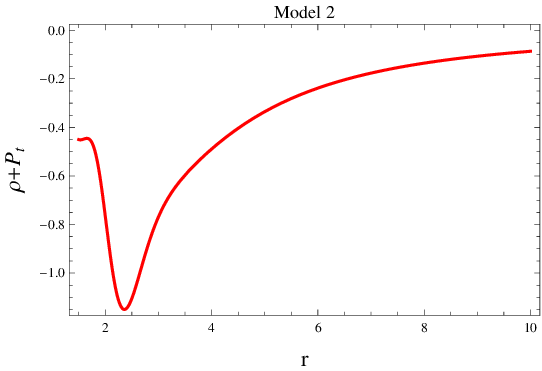,width=.5\linewidth}
\epsfig{file=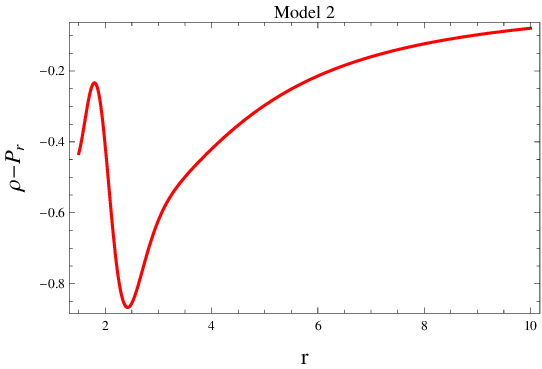,width=.5\linewidth}\epsfig{file=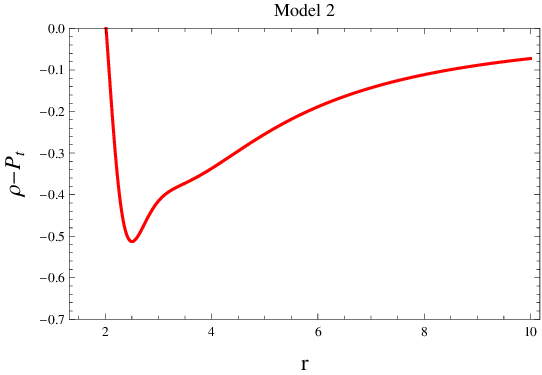,width=.5\linewidth}
\epsfig{file=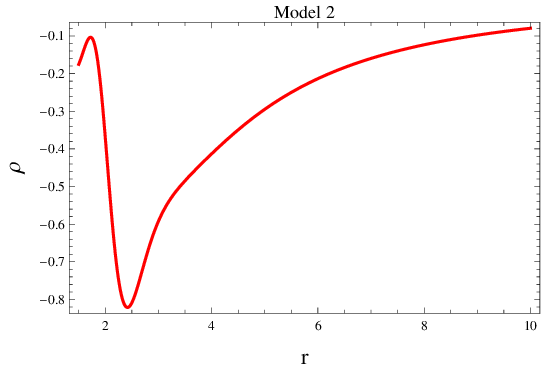,width=.5\linewidth}\epsfig{file=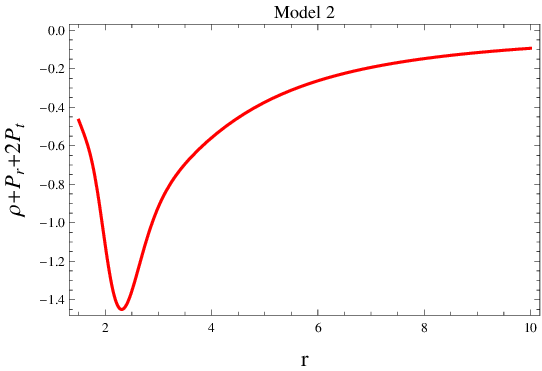,width=.5\linewidth}
\caption{\label{F11}Graphs of energy bounds for $a_{2}=2$,
$b_{2}=5$, $n_{2}=3$, $m=2$ and $\gamma=0.001$.}
\end{figure}

\begin{figure}
\epsfig{file=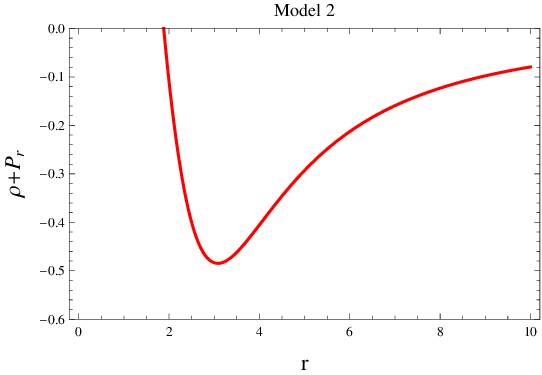,width=.5\linewidth}\epsfig{file=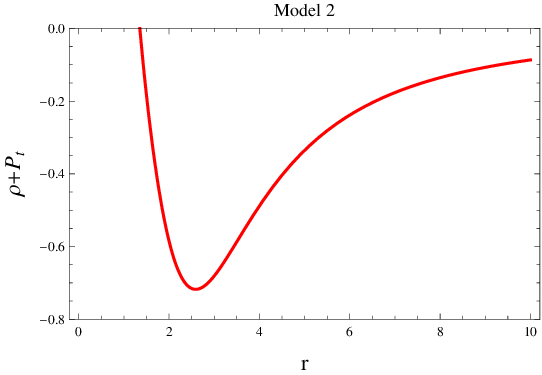,width=.5\linewidth}
\epsfig{file=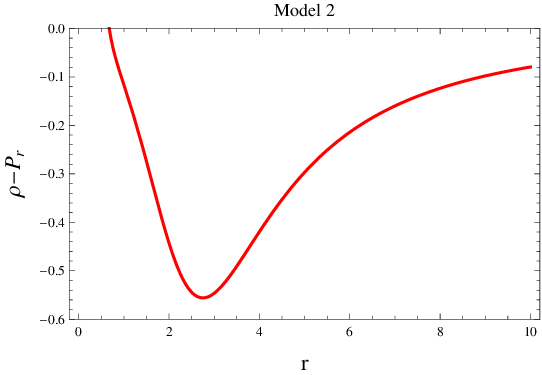,width=.5\linewidth}\epsfig{file=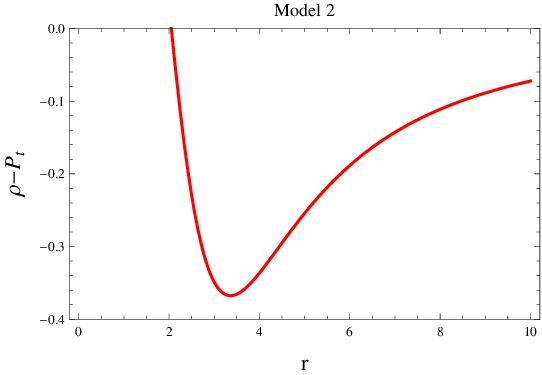,width=.5\linewidth}
\epsfig{file=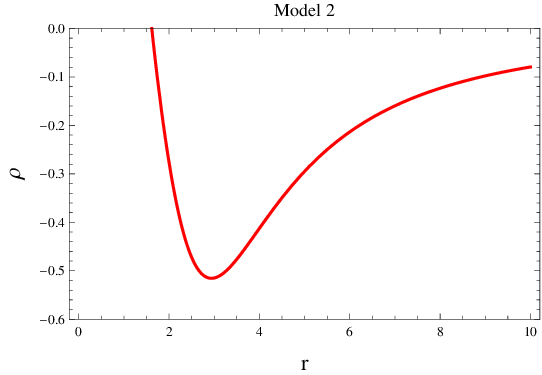,width=.5\linewidth}\epsfig{file=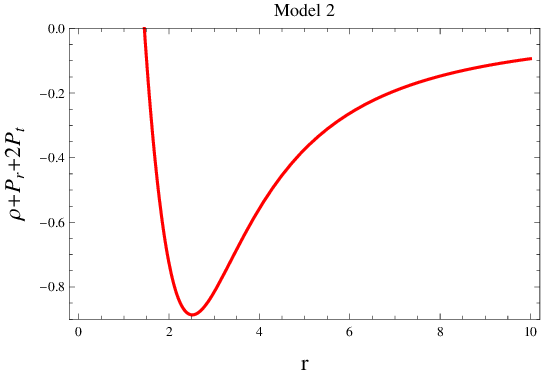,width=.5\linewidth}
\caption{\label{F12}Graphs of energy bounds for $a_{2}=-0.002$,
$b_{2}=-0.0003$, $n_{2}=1$, $m=-0.0003$ and $\gamma=0.001$.}
\end{figure}

\begin{figure}
\epsfig{file=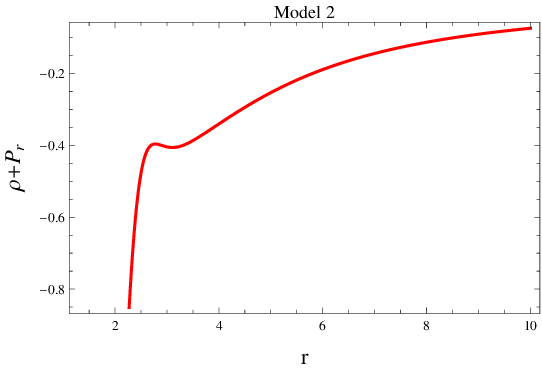,width=.5\linewidth}\epsfig{file=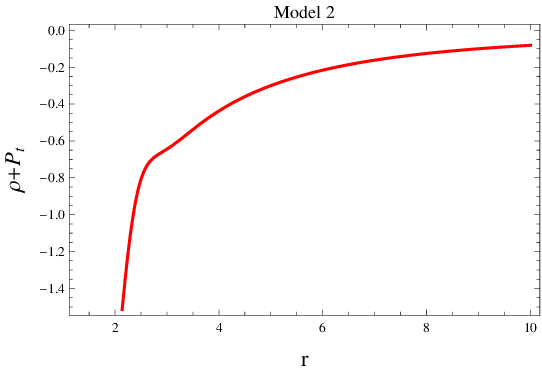,width=.5\linewidth}
\epsfig{file=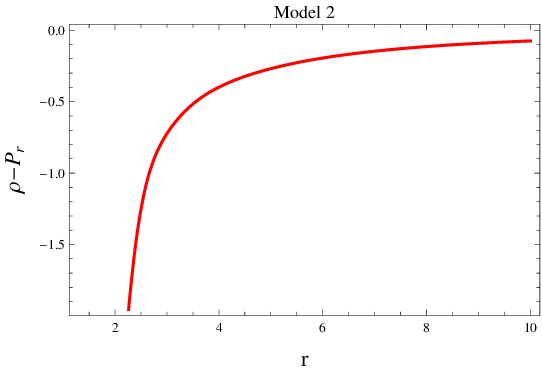,width=.5\linewidth}\epsfig{file=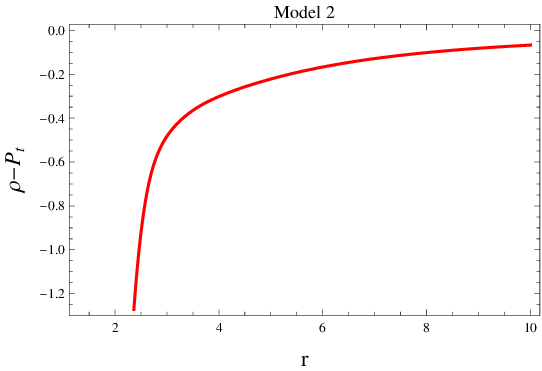,width=.5\linewidth}
\epsfig{file=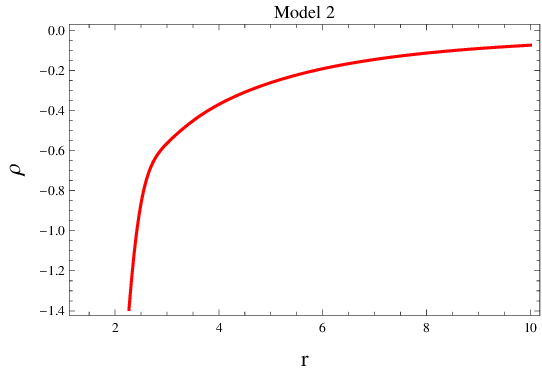,width=.5\linewidth}\epsfig{file=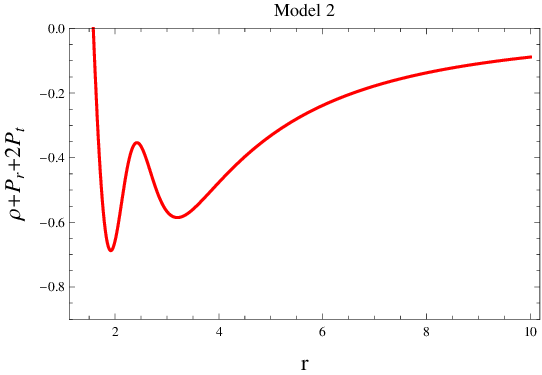,width=.5\linewidth}
\caption{\label{F13}Graphs of energy bounds for $a_{2}=-2$,
$b_{2}=-5$, $n_{2}=3$, $m=-2$ and $\gamma=0.001$.}
\end{figure}

Here, we use another model as \cite{41}
\begin{equation}\nonumber
f(\mathcal{G},T)=a_{2}\mathcal{G}^{n_{2}}(b_{2}\mathcal{G}^{m}+1)+\gamma
T,
\end{equation}
where $a_{2}$, $b_{2}$ and $m$ are arbitrary constant and $n_{2}>0$.
Appendix \textbf{B} contains the corresponding field equations. The
graphical behavior of null energy condition for different values of
$a_{2}$, $b_{2}$, $m$ and $n_{2}$ is given in Figures
\ref{F10}-\ref{F13}. These plots show that the null energy condition
violates for all values of the model parameter which gives the
viable traversable $\mathrm{WH}$ geometry.

\subsection{Model 3}

\begin{figure}
\epsfig{file=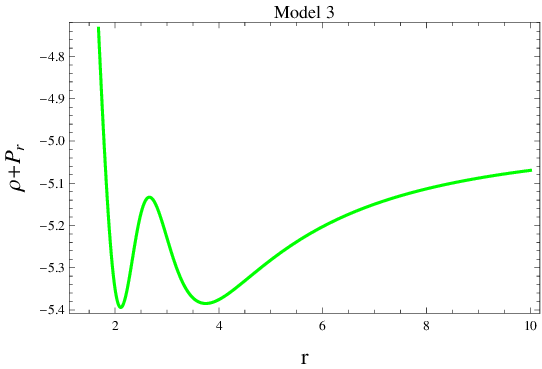,width=.5\linewidth}\epsfig{file=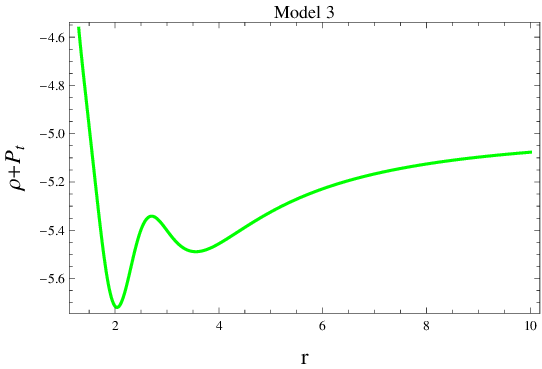,width=.5\linewidth}
\epsfig{file=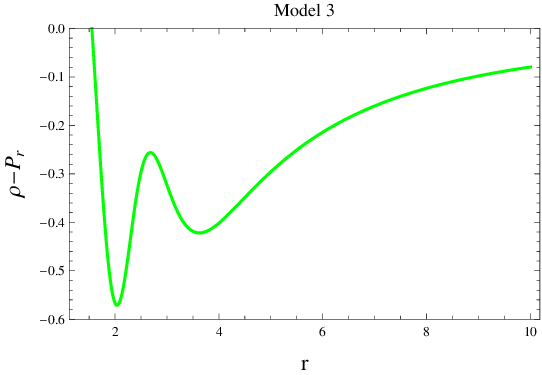,width=.5\linewidth}\epsfig{file=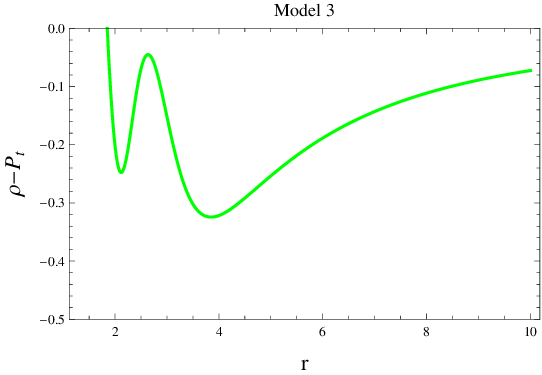,width=.5\linewidth}
\epsfig{file=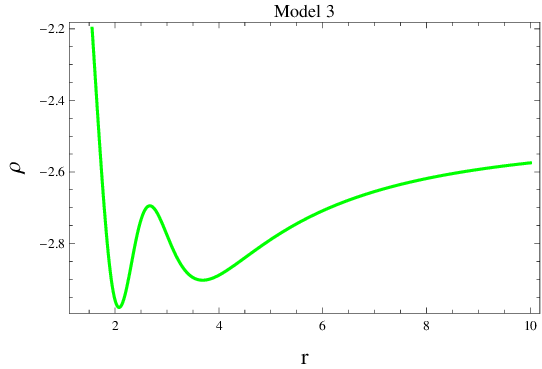,width=.5\linewidth}\epsfig{file=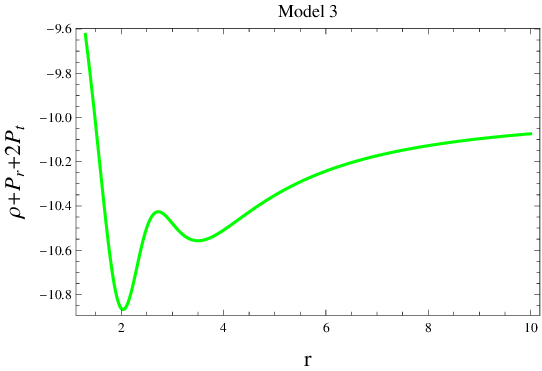,width=.5\linewidth}
\caption{\label{F14}Graphs of energy bounds for $a_{3}=0.01$,
$b_{3}=0.02$, $a_{4}=0.003$, $b_{4}=0.004$, $n_{3}=2$ and
$\gamma=0.001$.}
\end{figure}

\begin{figure}
\epsfig{file=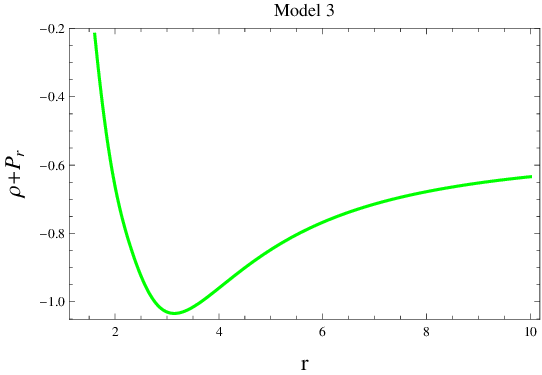,width=.5\linewidth}\epsfig{file=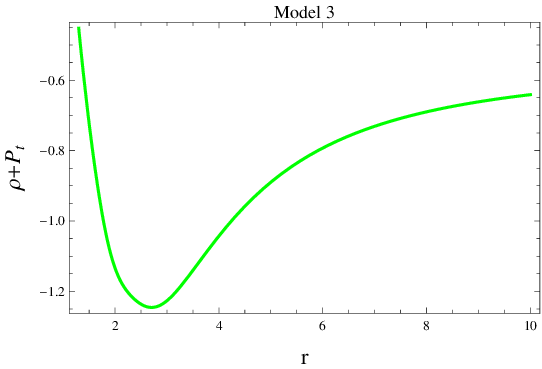,width=.5\linewidth}
\epsfig{file=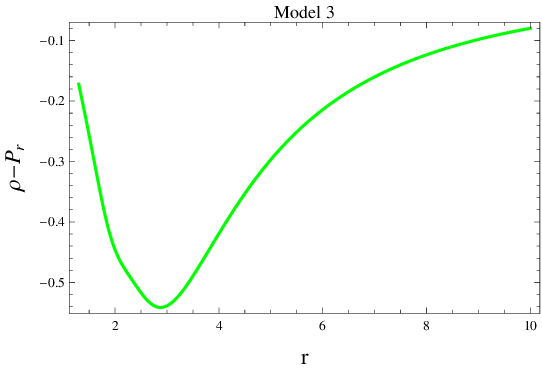,width=.5\linewidth}\epsfig{file=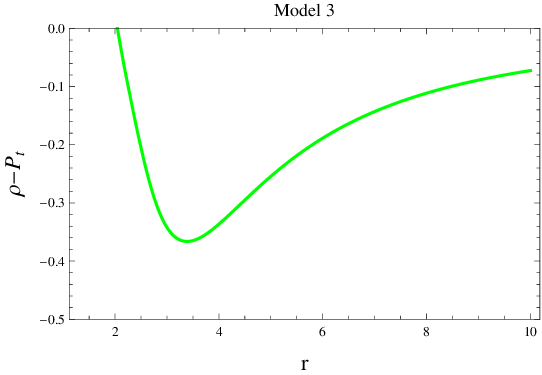,width=.5\linewidth}
\epsfig{file=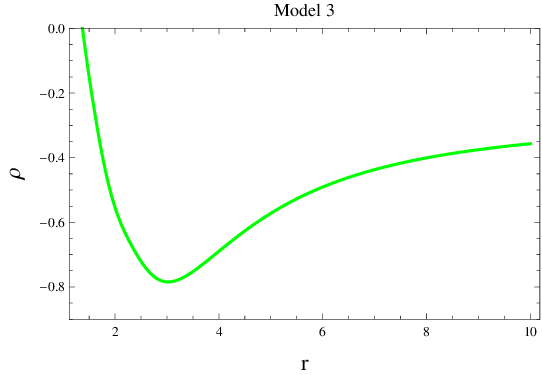,width=.5\linewidth}\epsfig{file=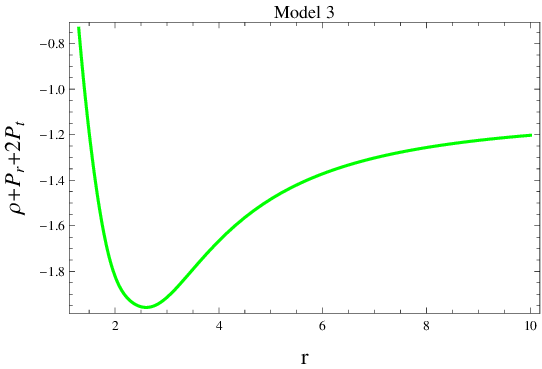,width=.5\linewidth}
\caption{\label{F15}Graphs of energy bounds for $a_{3}=2$,
$b_{3}=5$, $a_{4}=7$, $b_{4}=9$, $n_{3}=3$ and $\gamma=0.001$.}
\end{figure}

\begin{figure}
\epsfig{file=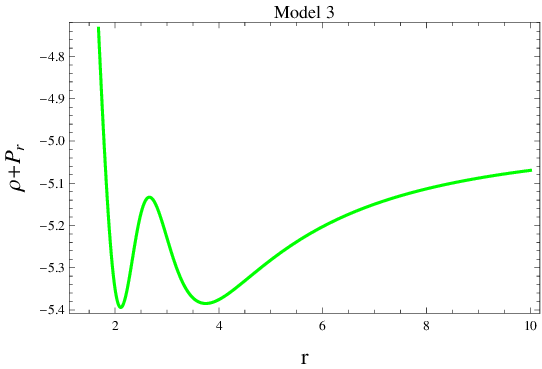,width=.5\linewidth}\epsfig{file=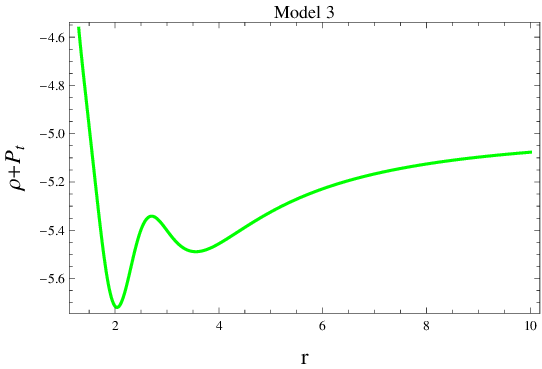,width=.5\linewidth}
\epsfig{file=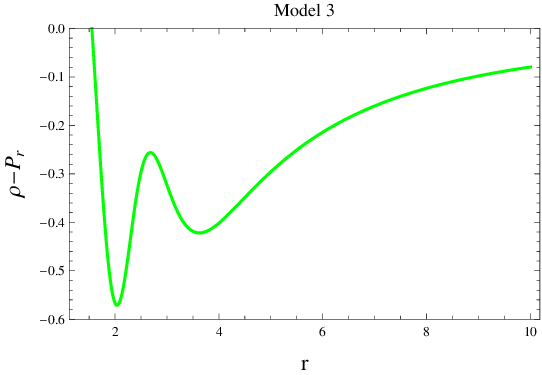,width=.5\linewidth}\epsfig{file=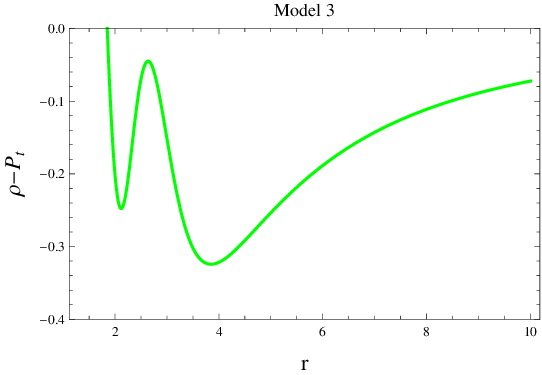,width=.5\linewidth}
\epsfig{file=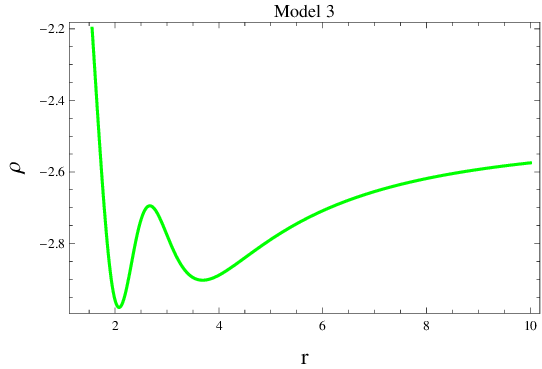,width=.5\linewidth}\epsfig{file=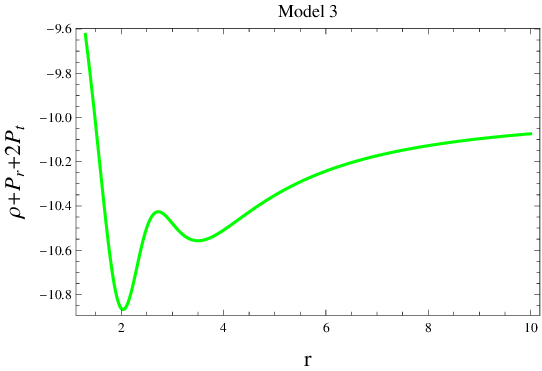,width=.5\linewidth}
\caption{\label{F16}Graphs of energy bounds for $a_{3}=-0.01$,
$b_{3}=-0.02$, $a_{4}=-0.003$, $b_{4}=-0.004$, $n_{3}=2$ and
$\gamma=0.001$.}
\end{figure}

\begin{figure}
\epsfig{file=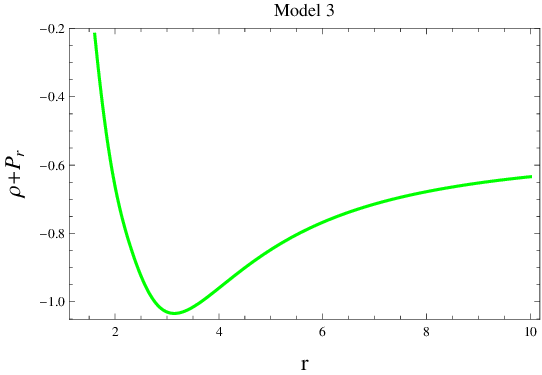,width=.5\linewidth}\epsfig{file=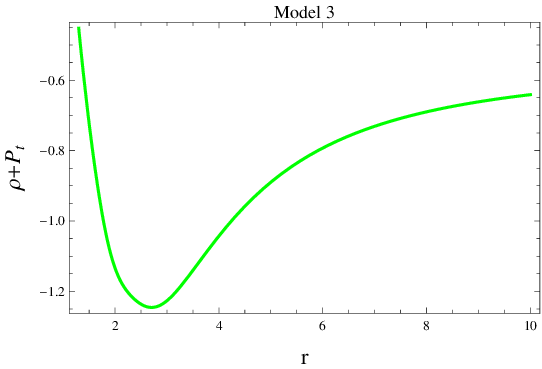,width=.5\linewidth}
\epsfig{file=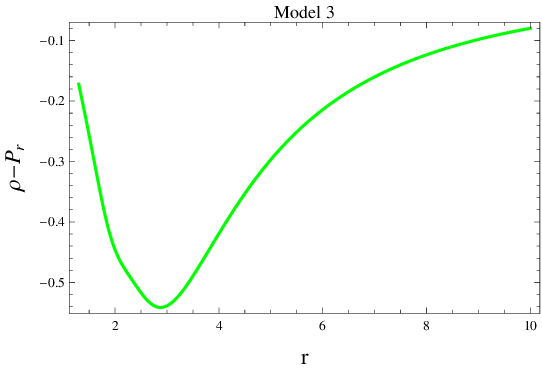,width=.5\linewidth}\epsfig{file=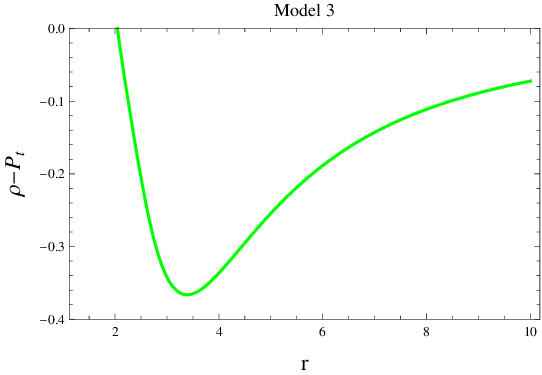,width=.5\linewidth}
\epsfig{file=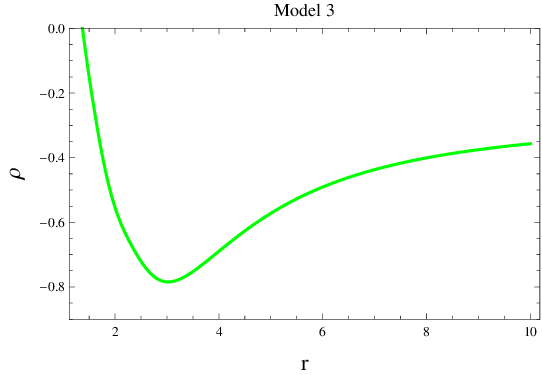,width=.5\linewidth}\epsfig{file=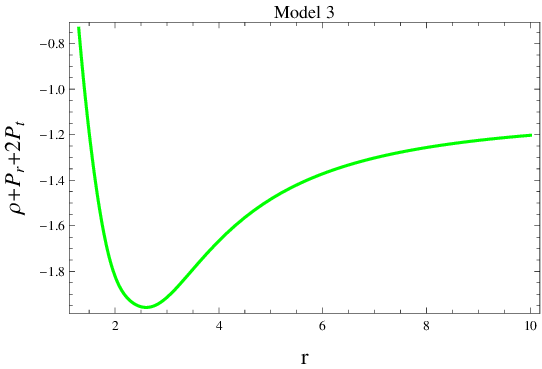,width=.5\linewidth}
\caption{\label{F17}Graphs of energy bounds for $a_{3}=-2$,
$b_{3}=-5$, $a_{4}=-7$, $b_{4}=-9$, $n_{3}=3$ and $\gamma=0.001$.}
\end{figure}

Finally, we consider the viable $f(\mathcal{G},T)$ model as
\begin{equation}\nonumber
f(\mathcal{G},T)=\frac{a_3\mathcal{G}^{n_{3}}+b_3}{a_4\mathcal{G}^{n_{3}}+b_4}+\gamma
T.
\end{equation}
here $a_{3}$, $a_{4}$, $b_{3}$, $b_{4}$ and $n_{3}$ are arbitrary
constants with $n_{3}>0$. Appendix \textbf{C} represents the field
equations corresponding to this model. Figures \ref{F14}-\ref{F17}
show the graphical behavior of energy conditions for various values
of the model parameters. These plots indicate the presence of viable
traversable $\mathrm{WH}$ for all parametric values.

\section{Stability Analysis}

The analysis of stability plays a crucial role in evaluating the
integrity of cosmic structures, particularly in the context of a
traversable $\mathrm{WH}$. This analysis involves investigating how
matter behaves within the $\mathrm{WH}$ and assessing its ability to
maintain structural integrity, preventing collapse. Examining the
stability of a $\mathrm{WH}$ entails utilizing field equations that
describe spacetime curvature in the presence of matter and energy.
According to quantum mechanics, particles can spontaneously appear
and vanish in space, causing fluctuations in the energy density of
the vacuum. These fluctuations, depending on their properties, can
either stabilize or destabilize the $\mathrm{WH}$. The notion of a
stable $\mathrm{WH}$ holds the potential to deepen our understanding
of the fundamental nature of spacetime and may lead to advancements
in technology and new forms of space travel.

\subsection{Causality Condition}

\begin{figure}
\epsfig{file=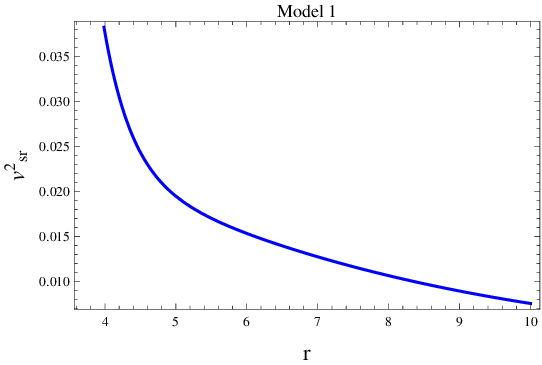,width=.5\linewidth}\epsfig{file=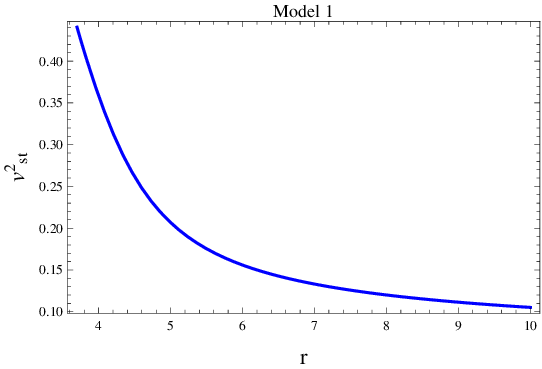,width=.5\linewidth}
\caption{\label{F18}Behavior of $v^{2}_{sr}$ and $v^{2}_{st}$ for
$a_{1}=1$, $b_{1}=0.1$, $n_{1}=2$ and $\gamma=0.001$.}
\end{figure}

\begin{figure}
\epsfig{file=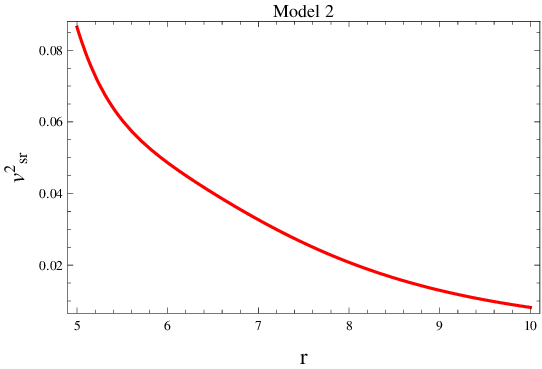,width=.5\linewidth}\epsfig{file=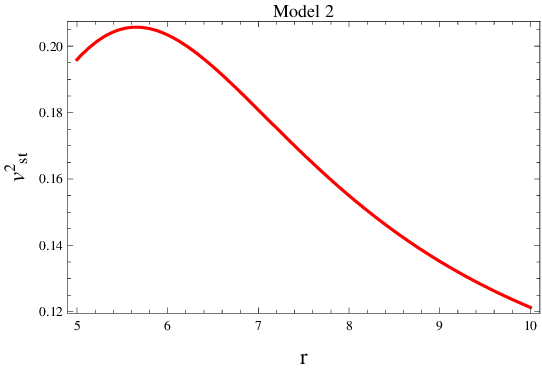,width=.5\linewidth}
\caption{\label{F19}Behavior of $v^{2}_{sr}$ and $v^{2}_{st}$ for
$a_{2}=2$, $b_{2}=5$, $n_{2}=2$, $m=0.3$ and $\gamma=0.001$.}
\end{figure}

\begin{figure}
\epsfig{file=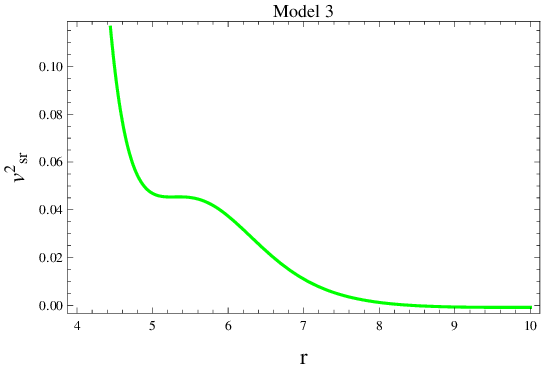,width=.5\linewidth}\epsfig{file=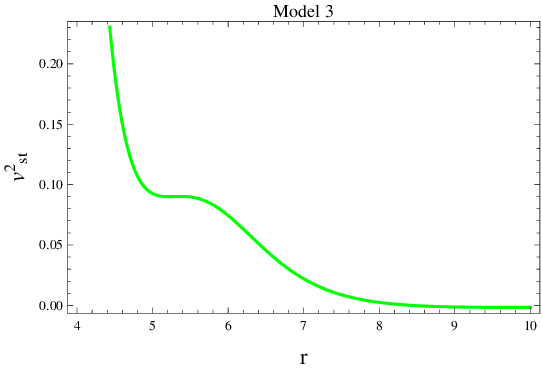,width=.5\linewidth}
\caption{\label{F20}Behavior of $v^{2}_{sr}$ and $v^{2}_{st}$ for
$a_{3}=-2$, $b_{3}=-5$, $a_{4}=0.7$ $b_{4}=0.09$, $n_{3}=2$ and
$\gamma=0.001$.}
\end{figure}

This criterion is a crucial element in the stability analysis in the
realm of gravitational physics. The application of this criterion
allows for the assessment of whether specific physical processes can
transmit signals at speeds surpassing that of light. Permitting such
faster-than-light propagation would result in causality violations,
giving rise to paradoxes and inconsistencies within the system. The
condition articulates that the components of sound speed, defined as
\begin{eqnarray}\nonumber
v^{2}_{sr}=\frac{dP_{r}}{d\rho}, \quad v_{st}^{2}
=\frac{dP_{t}}{d\rho},
\end{eqnarray}
must lie in [0,1] for stable structures. Figures \ref{F18}-\ref{F20}
show that the $\mathrm{WH}$ solutions are in the stable state as the
necessary conditions are satisfied.

\subsection{Herrera Cracking Technique}

\begin{figure}
\epsfig{file=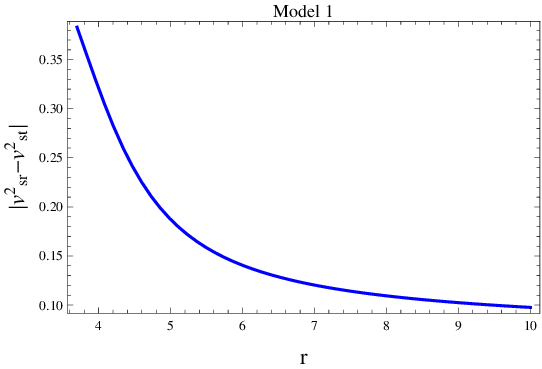,width=.5\linewidth}
\caption{\label{F21}Behavior of $|v^{2}_{sr}-v^{2}_{st}|$ for for
$a_{1}=1$, $b_{1}=0.1$, $n_{1}=2$ and $\gamma=0.001$.}
\end{figure}

\begin{figure}
\epsfig{file=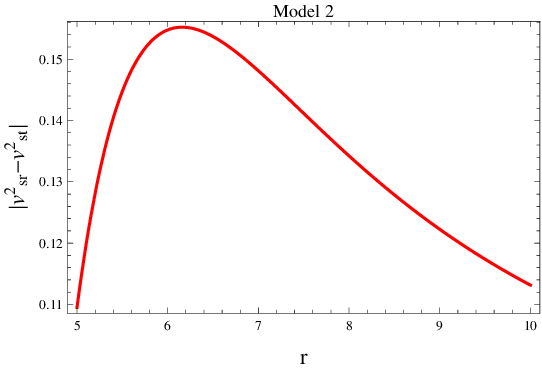,width=.5\linewidth} \caption{\label{F22}
Behavior of $|v^{2}_{sr}-v^{2}_{st}|$ for $a_{2}=2$, $b_{2}=5$,
$n_{2}=2$, $m=0.3$ and $\gamma=0.001$.}
\end{figure}

\begin{figure}
\epsfig{file=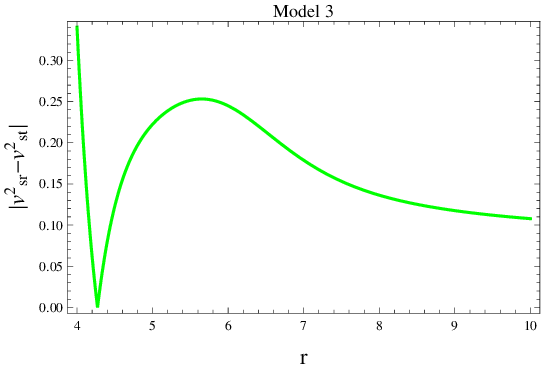,width=.5\linewidth}
\caption{\label{F23}Behavior of $|v^{2}_{sr}-v^{2}_{st}|$ for
$a_{3}=-2$, $b_{3}=-5$, $a_{4}=0.7$ $b_{4}=0.09$, $n_{3}=2$ and
$\gamma=0.001$.}
\end{figure}

The stability analysis of solutions employs a mathematical method,
as introduced by Herrera \cite{42}, known as the cracking approach.
This method provides a theoretical framework for examining the
stability of cosmic systems. According to this approach, the fluid
within these systems may undergo a "cracking" or breaking process
due to the gravitational and pressure conflicts. The key premise is
that a stable cosmic system must satisfy specific requirements
related to the speed of sound. In particular, the difference in
sound speed components should fall within the range of 0 to 1 for
stability. Failure to meet this condition indicates instability,
potentially leading to system collapse. Utilizing this technique
enables researchers to assess the stability of $\mathrm{WH}$
solutions, offering crucial insights into the behavior of these
exotic structures in the universe. The graphical representation of
$|v^{2}{sr}-v{st}^{2}|$ in Figures \ref{F21}-\ref{F23} illustrate
that $\mathrm{WH}$s maintain stability even in the presence of
modified terms.

\subsection{Adiabatic Index}

This is another valuable technique employed for assessing the
stability of astronomical objects. This method is instrumental in
examining the stability of celestial bodies and gaining insights
into the characteristics of the substances within them. Ensuring the
stability of these entities is crucial, as any disturbance could
lead to collapse or explosion. The adiabatic index offers insights
into how matter reacts to alterations in pressure and density,
aiding in the determination of the stability of a $\mathrm{WH}$. The
components of the adiabatic index are defined as
\begin{eqnarray}\label{21}
\Gamma_{r}=\frac{\rho+P_{r}}{P_{r}}v^{2}_{sr},\quad
\Gamma_{t}=\frac{\rho+P_{t}}{P_{t}}v^{2}_{st}.
\end{eqnarray}

To assess the stability of a wormhole using the adiabatic index
method, one must determine the value of $\Gamma$ for the matter
composing the wormhole. A stable wormhole is characterized by a
value of $\Gamma$ greater than 4/3, while an unstable one will
collapse if the value of   $\Gamma$ is less than 4/3. This
instability arises from the susceptibility of the material to mix
under small perturbations, leading to energy loss and collapse.
Figures 24-26 illustrate the stability of our system in correlation
with models \textbf{1} and \textbf{2}, both displaying stability.
However, for model \textbf{3}, the adiabatic index is less than 4/3,
indicating instability. Consequently, viable traversable stable
$\mathrm{WH}$ solutions are only obtained for models \textbf{2} and
\textbf{3}.

\section{Final Remarks}

In this manuscript, we have investigated the feasibility of
traversable $\mathrm{WH}$ solutions using the Karmarkar technique in
the framework of $f(\mathcal{G}, T)$ theory. Our primary aim is to
examine the theoretical aspects of $\mathrm{WH}$ solutions and their
behavior under modified terms. Traversable $\mathrm{WH}$s have
garnered significant attention in theoretical physics due to their
potential to connect different regions of spacetime. Identifying
viable and stable traversable WH solutions in $f(\mathcal{G},T)$
theory is not just a theoretical exercise rather, it holds profound
implications for our understanding of the fundamental laws governing
the universe. Firstly, our findings suggest that traversable
$\mathrm{WH}$s can exist in modified gravity theories such as
$f(\mathcal{G},T)$ theory. This challenges the concept that such
structures are only theoretically feasible in the framework of GTR.
By demonstrating their existence in this modified framework, we have
expanded the scope for studying and investigating these intriguing
phenomena with a wider range of theoretical frameworks than
previously considered. Secondly, the stability of the $\mathrm{WH}$
solutions are crucial for their practical implications. Our analysis
indicates that these solutions satisfy causality conditions and do
not exhibit instability akin to Herrera cracking. This implies that
they could serve as stable pathways for traversing between distant
regions of the universe. Thus, our research advances the
understanding of traversable WHs in a modified framework and lays a
theoretical foundation for further exploration.

This manuscript delves into the viability of traversable
$\mathrm{WH}$ configurations in the framework of $f(\mathcal{G},T)$
theory. The primary objective is to examine the behavior of the
shape function and energy conditions in this context. Wormholes, as
solutions to Einstein's field equations have gained considerable
interest for their implications in cosmology and interstellar
travel. However, their viability and stability under alternative
gravitational theories remain uncertain. The motivation behind
exploring $f(\mathcal{G}, T)$ gravity is twofold. Firstly, this
gravitational theory extends GTR to provide a more comprehensive
understanding of gravitational phenomena. Secondly, $\mathrm{WH}$
solutions in $f(\mathcal{G}, T)$ gravity offer novel insights into
the relationship between gravity modifications and exotic
structures. Through the investigation of $\mathrm{WH}$s in this
modified gravity framework, we aim to enhance our comprehension of
their existence and stability. Additionally, this exploration may
illuminate the compatibility of $\mathrm{WH}$s with modified gravity
theories, thereby impacting theoretical physics and observational
cosmology.
\begin{figure}
\epsfig{file=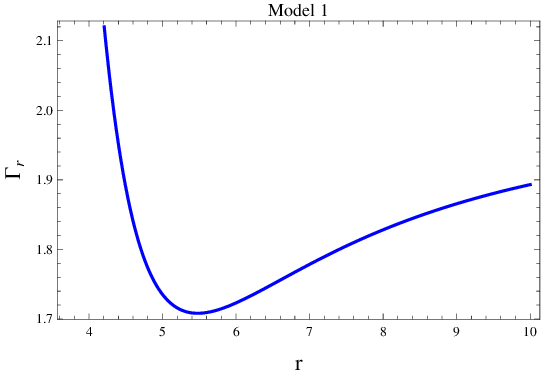,width=.5\linewidth}\epsfig{file=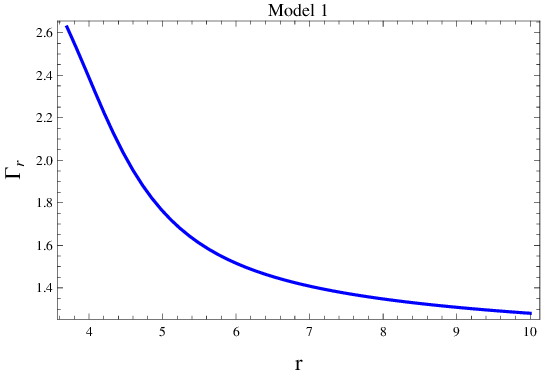,width=.5\linewidth}
\caption{\label{F24}Behavior of $\Gamma_{r}$ and $\Gamma_{t}$ for
$a_{1}=1$, $b_{1}=0.1$, $n_{1}=2$ and $\gamma=0.001$.}
\end{figure}

The shape function obtained through the Karmarkar condition leads to
a viable $\mathrm{WH}$ geometry, meeting all essential conditions.
In the case of a power law model, violation of the null energy
condition occurs for certain parameter values, indicating the
presence of exotic matter at the $\mathrm{WH}$ throat and thereby
confirming the existence of a viable traversable $\mathrm{WH}$.
Conversely, when the null energy condition is satisfied, no exotic
matter is present, and a viable traversable $\mathrm{WH}$ is not
attained. $\mathrm{WH}$ solutions exist for all parameter values in
the second model, indicating the presence of a viable traversable
$\mathrm{WH}$ geometry. The third model confirms the existence of
exotic matter at the $\mathrm{WH}$ throat. Stability considerations,
including causality conditions, Herrera cracking, and the adiabatic
index, were satisfied for the first two models, ensuring the
existence of physically viable and stable $\mathrm{WH}$ geometries.
However, for the third model, the adiabatic components' values being
less than 4/3 suggest the presence of an unstable $\mathrm{WH}$.

\begin{figure}
\epsfig{file=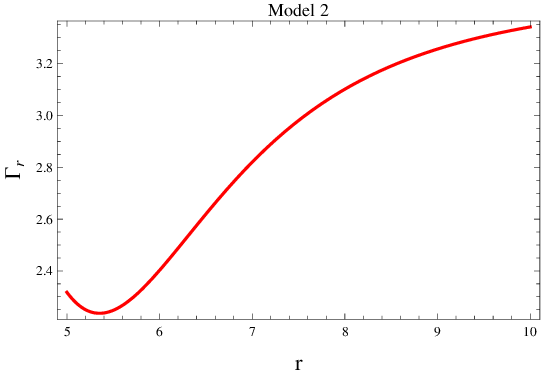,width=.5\linewidth}\epsfig{file=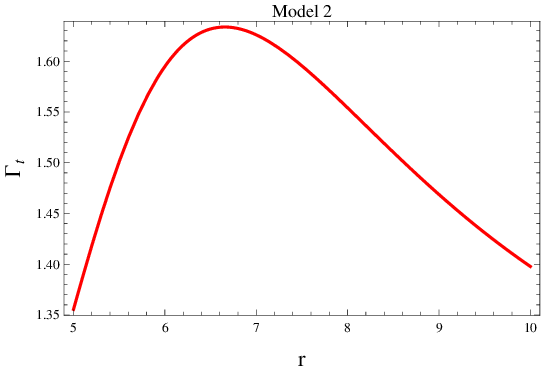,width=.5\linewidth}
\caption{\label{F25}Behavior of $\Gamma_{r}$ and $\Gamma_{t}$ for
$a_{2}=2$, $b_{2}=5$, $n_{2}=2$, $m=0.3$ and $\gamma=0.001$.}
\end{figure}
Traversable $\mathrm{WH}$s represent a captivating subject in
theoretical physics, particularly in the framework of
$f(\mathcal{G}, T)$ theory. The exploration of wormholes holds
significant relevance and theoretical significance. Their existence
introduces the intriguing prospect of time travel by enabling the
formation of closed timelike curves. In the context of
$f(\mathcal{G}, T)$ gravitational theory, the presence of
traversable wormholes challenges conventional notions of spacetime
geometry. It implies that deviations from the standard
Einstein-Hilbert action can yield exotic structures like wormholes,
thereby opening up new avenues for investigating the foundational
principles of physics. The analysis of traversable wormholes offers
valuable insights into the impact of modified gravity on the
large-scale structure of the universe. This contributes to refining
our models of cosmic evolution and understanding the eventual fate
of the universe. Moreover, the existence of viable traversable
wormholes in $f(\mathcal{G}, T)$ theory facilitates a deeper
exploration of the energy conditions governing spacetime and their
compatibility with exotic matter. Consequently, the practical
implications of traversable wormholes in $f(\mathcal{G}, T)$ theory
have the potential to revolutionize both space travel and cosmology,
while their theoretical ramifications reshape our comprehension of
fundamental physics, encompassing gravity, spacetime and the essence
of the cosmos.

Fayyaz and Shamir \cite{43} investigated viable traversable
$\mathrm{WH}$ geometries using Karmarkar condition in the framework
of GTR. They developed a suitable shape function through the
Karmarkar technique which satisfies necessary Morris-Throne
conditions. Their analysis yielded viable $\mathrm{WH}$ structures,
although stability analysis was not conducted. Later, they extended
their investigation to $f(\mathfrak{R})$ theory, where they achieved
viable $\mathrm{WH}$ solutions with minimal exotic matter \cite{33}.
However, stability analysis was not performed in this modified
framework. Recently, Sharif and Fatima \cite{44} extended this line
of research to $f(\mathfrak{R}, T)$ theory, where they successfully
derived viable and stable $\mathrm{WH}$ solutions for the minimum
radius. Furthermore, we have identified the viable traversable
$\mathrm{WH}$ solutions in the context of $f(\mathcal{G}, T)$
gravity as the energy conditions are violated which indicates the
presence of exotic matter at $\mathrm{WH}$ throat. This implies that
the physically viable traversable $\mathrm{WH}$ solutions exist in
$f(\mathcal{G}, T)$ theory.
\begin{figure}
\epsfig{file=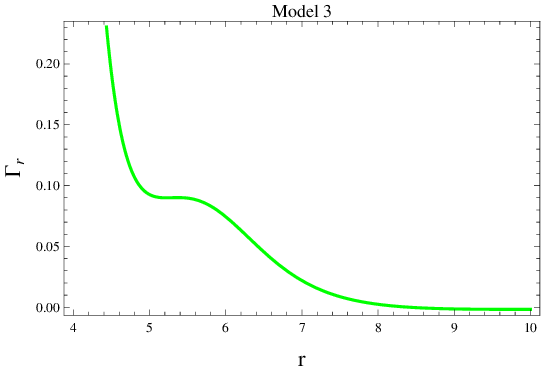,width=.5\linewidth}\epsfig{file=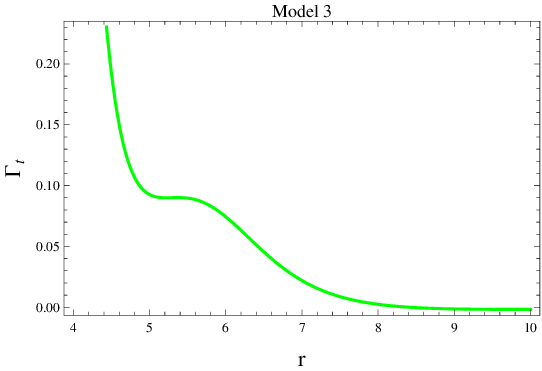,width=.5\linewidth}
\caption{\label{F26}Behavior of $\Gamma_{r}$ and $\Gamma_{t}$ for
$a_{3}=-2$, $b_{3}=-5$, $a_{4}=0.7$ $b_{4}=0.09$, $n_{3}=2$ and
$\gamma=0.001$.}
\end{figure}

\vspace{0.25cm}

\section*{Appendix A}
\renewcommand{\theequation}{A\arabic{equation}}
\setcounter{equation}{0}

The corresponding field equations for model \textbf{1} are
\begin{eqnarray}\nonumber
\rho&=&\frac{e^{-2\eta}}{8r^{4}(1+\gamma)(1+2\gamma)}
\bigg[8e^{\eta}(e^{\eta}-1)r^2(1+2\gamma)
-4r^{4}e^{2\eta}r^{4}(1+\gamma)(a_{1}
\mathcal{G}^{n_{1}}+b_{1}\ln(\mathcal{G}) \mathcal{G})+(a_{1}
n_{1}\mathcal{G}^{n_{1}-1}+b_{1}+\ln(\mathcal{G})b_{1})
\\\nonumber
&\times&\bigg\{-16(e^{\eta}-1)^{2}\gamma+r^{2}
\big\{r^{2}(1+2\gamma)\xi'^{4}
-2r^{2}(1+2\gamma)\xi'^{3}\eta'-4\xi'\eta'\big\{2(e^{\eta}-3)(1+\gamma)
+r^{2}(1+2\gamma)\xi''+\xi'^{2}(8(\gamma
\\\nonumber
&+&(1+\gamma)e^{\eta}))+r^{2}(1+2\gamma)(\eta'^{2}+4\xi'')\big\}
+4\big\{-2\gamma
\eta'^{2}+\xi''(4(e^{\eta}-1)(1+\gamma))\big\}+r^{2}(1+2\gamma)\xi''\big\}\bigg\}
+2r\bigg\{4(a_{1} n_{1}
\\\nonumber
&\times&(n_{1}-1)\mathcal{G}^{n_{1}-2}\mathcal{G}'+\frac{b_{1}}
{\mathcal{G}}\mathcal{G}')\big\{-8(2+5\gamma)+
r(\eta'(10+27\gamma-2r\gamma \eta'))
-r(8+18\gamma+r(2+3\gamma)\eta')\xi''\big\}-8r(2+5\gamma)
\\\nonumber
&\times&(1-2r\eta'+r^{2}\xi'')\big\{a_{1}
n_{1}(n_{1}-1)(n_{1}-2)\mathcal{G}^{n_{1}-3}\mathcal{G}'^{2}+a_{1}
n_{1}(n_{1}-1)\mathcal{G}^{n_{1}-2}\mathcal{G}''+\frac{b_{1}}{\mathcal{G}^{2}}
(\mathcal{G}\mathcal{G}''-\mathcal{G}'^{2})\big\}
+r^{2}\xi'^{2}\bigg\{re^{\eta}\gamma
\\\nonumber
&-&2(8+18\gamma+r(2+3\gamma)\eta')(a_{1}
n_{1}(n_{1}-1)\mathcal{G}^{n_{1}-2}\mathcal{G}'+\frac{b_{1}}
{\mathcal{G}}\mathcal{G}')-4r(2+5\gamma)\big\{a_{1}
n_{1}(n_{1}-1)(n_{1}-2)\mathcal{G}^{n_{1}-3}\mathcal{G}'^{2}+a_{1}
n_{1}
\\\nonumber
&\times&(n_{1}-1)\mathcal{G}^{n_{1}-2}\mathcal{G}''+\frac{b_{1}}
{\mathcal{G}^{2}}(\mathcal{G}\mathcal{G}''-\mathcal{G}'^{2})\big\}\bigg\}
+2e^{\eta}\bigg\{16(2+5\gamma)(a_{1}
n_{1}(n_{1}-1)\mathcal{G}^{n_{1}-2}\mathcal{G}'+\frac{b_{1}}
{\mathcal{G}}\mathcal{G}')+2r\eta'\big\{r+2r\gamma
\\\nonumber
&+&(2+3\gamma)(a_{1}
n_{1}(n_{1}-1)\mathcal{G}^{n_{1}-2}\mathcal{G}'+\frac{b_{1}}{\mathcal{G}}
\mathcal{G}')\big\} +r^{3}\gamma\xi''+4r(2+5\gamma)\big\{a_{1}
n_{1}(n_{1}-1)(n_{1}-2)\mathcal{G}^{n_{1}-3}\mathcal{G}'^{2}+a_{1}
n_{1}(n_{1}-1)
\\\nonumber
&\times&\mathcal{G}^{n_{1}-2}\mathcal{G}''+\frac{b_{1}}{\mathcal{G}^{2}}
(\mathcal{G}\mathcal{G}''-\mathcal{G}'^{2})\big\}\bigg\}+ra'\big\{2(-32
-74\gamma+r\eta'(2\gamma+r(2+3\gamma)\eta'))\big\}(a_{1}
n_{1}(n_{1}-1)\mathcal{G}^{n_{1}-2}\mathcal{G}'+\frac{b_{1}}{\mathcal{G}}
\mathcal{G}')
\\\nonumber
&-&e^{\eta}\gamma\big\{r(-4+r\eta')+4(a_{1}
n_{1}(n_{1}-1)\mathcal{G}^{n_{1}-2}\mathcal{G}'+\frac{b_{1}}
{\mathcal{G}}\mathcal{G}')\big\}
+4r\big\{-2(4+9\gamma)+r(2+5\gamma)\eta'\big\{a_{1}
n_{1}(n_{1}-1)(n_{1}-2)
\\\label{21}
&\times&\mathcal{G}^{n_{1}-3}\mathcal{G}'^{2}+a_{1}
n_{1}(n_{1}-1)\mathcal{G}^{n_{1}-2}\mathcal{G}''+\frac{b_{1}}{\mathcal{G}^{2}}
(\mathcal{G}\mathcal{G}''-\mathcal{G}'^{2})\big\}\big\}\bigg\}\bigg],
\\\nonumber
p_{r}&=&\frac{e^{-2\eta}}{8r^{4}(1+\gamma)(1+2\gamma)}
\bigg[-4e^{2\eta}r^{4}(1+\gamma)(a_{1}
\mathcal{G}^{n_{1}}+b_{1}\ln(\mathcal{G})\mathcal{G})+(a_{1} n_{1}
\mathcal{G}^{n_{1}-1}+b_{1}+\ln(\mathcal{G})b_{1})\bigg\{-16(e^{\eta}-1)^{2}\gamma
\\\nonumber
&+&r^{2}\bigg\{r^{2}(1+2\gamma)\xi'^{4}
-2r^{2}(1+2\gamma)\xi'^{3}\eta'-4\xi'\eta'(2(e^{\eta}-3)(1+\gamma)
+r^{2}(1+2\gamma)\xi'')+\xi'^{2}\big\{8(e^{\eta}-1)(1+\gamma)
\\\nonumber
&+&r^{2}(1+2\gamma)(\eta'^{2}+4\xi'')\big\}+4\big\{2(1+\gamma)
\eta'^{2}+\xi''(4(e^{\eta}-1)(1+\gamma)+r^{2}(1+2\gamma)\xi'')\big\}\bigg\}\bigg\}
+2r(4e^{\eta}(e^{\eta}-1))r
\\\nonumber
&\times&(1+2\gamma)+4(a_{1}
n_{1}(n_{1}-1)\mathcal{G}^{n_{1}-2}\mathcal{G}'+\frac{b_{1}}{\mathcal{G}}
\mathcal{G}')\bigg\{-8\gamma +r\bigg\{-\eta'(4+\gamma+2r\gamma
\eta')+r\gamma(-2+r \eta'\xi'')+r^{2}\gamma\xi'^{2}
\\\nonumber
&\times&\bigg\{2(-2+r\eta')(a_{1}
n_{1}(n_{1}-1)\mathcal{G}^{n_{1}-2}\mathcal{G}'+\frac{b_{1}}{\mathcal{G}}
\mathcal{G}')+r\big\{e^{\eta}-4\big(a_{1}
n_{1}(n_{1}-1)(n_{1}-2)\mathcal{G}^{n_{1}-3}\mathcal{G}'^{2}+a_{1}
n_{1}(n_{1}-1)
\\\nonumber
&\times&\mathcal{G}^{n_{1}-2}\mathcal{G}''+\frac{b_{1}}{\mathcal{G}^{2}}
(\mathcal{G}\mathcal{G}''-\mathcal{G}'^{2})\big)\big\}\bigg\}-8r\gamma
(1-2r\eta'+r^{2}\xi'')\big\{a_{1}
n_{1}(n_{1}-1)(n_{1}-2)\mathcal{G}^{n_{1}-3}\mathcal{G}'^{2}+a_{1}
n_{1}(n_{1}-1)
\\\nonumber
&\times&\mathcal{G}^{n_{1}-2}\mathcal{G}''+\frac{b_{1}}{\mathcal{G}^{2}}
(\mathcal{G}\mathcal{G}''-\mathcal{G}'^{2})\big\}
+ra'\bigg\{-2(12+34\gamma+r\eta'(8+14\gamma+4\gamma \eta'))(a_{1}
n_{1}(n_{1}-1)\mathcal{G}^{n_{1}-2}\mathcal{G}'
+\frac{b_{1}}{\mathcal{G}}\mathcal{G}')
\\\nonumber
&+&e^{\eta}\big\{-r(4+4\gamma+r\gamma \eta')+4(2+3\gamma)(a_{1}
n_{1}(n_{1}-1)\mathcal{G}^{n_{1}-2}\mathcal{G}'+\frac{b_{1}}
{\mathcal{G}}\mathcal{G}')\big\} +4r\gamma(-2+r\eta')\big\{a_{1}
n_{1}(n_{1}-1)(n_{1}-2)
\\\nonumber
&\times&\mathcal{G}^{n_{1}-3}\mathcal{G}'^{2}+a_{1}n_{1}(n_{1}-1)
\mathcal{G}^{n_{1}-2}\mathcal{G}''+\frac{b_{1}}{\mathcal{G}^{2}}
(\mathcal{G}\mathcal{G}''-\mathcal{G}'^{2})\big\}\bigg\}
+2e^{\eta}\big\{2(8\gamma+r(4+7\gamma)\eta')(a_{1}
n_{1}(n_{1}-1)\mathcal{G}^{n_{1}-2}\mathcal{G}'
\\\label{22}
&+&\frac{b_{1}}{\mathcal{G}}\mathcal{G}')+r\gamma\big\{r^2\xi''+4\big(a_{1}
n_{1}(n_{1}-1)(n_{1}-2)\mathcal{G}^{n_{1}-3}\mathcal{G}'^{2}+a_{1}
n_{1}(n_{1}-1)\mathcal{G}^{n_{1}-2}\mathcal{G}''+\frac{b_{1}}{\mathcal{G}^{2}}
(\mathcal{G}\mathcal{G}''-\mathcal{G}'^{2})\big)\big\}\big\}
\bigg\}\bigg\}\bigg],
\end{eqnarray}
\begin{eqnarray}\nonumber
p_{t}&=&\frac{e^{-2\eta}}{4r^{4}(1+\gamma)(1+2\gamma)}
\bigg[-2e^{2\eta}r^{4}(1+\gamma)(a_{1}
\mathcal{G}^{n_{1}}+b_{1}\ln(\mathcal{G})\mathcal{G}) +2(a_{1}
n_{1}\mathcal{G}^{n_{1}-1}+b_{1}+b_{1}
\ln(\mathcal{G}))\bigg\{4(e^{\eta}-1)^{2}(1+\gamma)
\\\nonumber
&+&r^{2}\big\{(-1+2e^{\eta}(1+\gamma))a'^{2}-2(-3+e^{\eta})
(1+\gamma)a'\eta'+\eta'^{2}+4(-1+e^{\eta})(1+\gamma)a''\big\}\bigg\}
-r\bigg\{4(a_{1} n_{1}(n_{1}-1)\mathcal{G}^{n_{1}-2}\mathcal{G}'
\\\nonumber
&+&\frac{b_{1}}{\mathcal{G}}\mathcal{G}')\big\{8\gamma+r(\eta'
(-7\gamma-2r(1+\gamma)\eta')
+r(2+6\gamma+r(2+3\gamma)\eta')a'')\big\}
+8r\gamma(1-2r\eta'+r^{2}a'')\big\{a_{1} n_{1}(n_{1}-1)
\\\nonumber
&\times&(n_{1}-2)\mathcal{G}^{n_{1}-3}\mathcal{G}'^{2}+a_{1}
n_{1}(n_{1}-1)\mathcal{G}^{n_{1}-2}\mathcal{G}''+\frac{b_{1}}{\mathcal{G}^{2}}
(\mathcal{G}\mathcal{G}''-\mathcal{G}'^{2})\big\}+2e^{\eta}\big\{-16\gamma
(a_{1}
n_{1}(n_{1}-1)\mathcal{G}^{n_{1}-2}\mathcal{G}'+\frac{b_{1}}{\mathcal{G}}\mathcal{G}')
\\\nonumber
&-&r\eta'(r+2r\gamma-2\gamma (a_{1}
n_{1}(n_{1}-1)\mathcal{G}^{n_{1}-2}\mathcal{G}'+\frac{b_{1}}{\mathcal{G}}
\mathcal{G}'))+r^{3}(1+\gamma)a''-4r\gamma \big\{a_{1}
n_{1}(n_{1}-1)(n_{1}-2)\mathcal{G}^{n_{1}-3}\mathcal{G}'^{2}+a_{1}
n_{1}
\\\nonumber
&\times&(n_{1}-1)
\mathcal{G}^{n_{1}-2}\mathcal{G}''+\frac{b_{1}}{\mathcal{G}^{2}}(\mathcal{G}
\mathcal{G}''-\mathcal{G}'^{2})\big\}
\big\}+r^{2}a'^{2}\bigg\{e^{\eta}r(1+\gamma)
+2(2+6\gamma+r(2+3\gamma)\eta')(a_{1}
n_{1}(n_{1}-1)\mathcal{G}^{n_{1}-2}\mathcal{G}'
\\\nonumber
&+&\frac{b_{1}}{\mathcal{G}}\mathcal{G}')+4r\gamma\big\{a_{1}
n_{1}(n_{1}-1)(n_{1}-2)\mathcal{G}^{n_{1}-3}\mathcal{G}'^{2}+a_{1}
n_{1}(n_{1}-1)\mathcal{G}^{n_{1}-2}\mathcal{G}''+\frac{b_{1}}{\mathcal{G}^{2}}
(\mathcal{G}\mathcal{G}''-\mathcal{G}'^{2})\big\}
\bigg\}+ra'-2(-10\gamma
\\\nonumber
&+&r\eta'(2+6\gamma+r(2+3\gamma)\eta'))(a_{1}
n_{1}(n_{1}-1)\mathcal{G}^{n_{1}-2}\mathcal{G}'+\frac{b_{1}}{\mathcal{G}}
\mathcal{G}')+e^{\eta}\big\{-r(-2+r(1+\gamma)\eta') +4\gamma(a_{1}
n_{1}(n_{1}-1)
\\\nonumber
&\times&\mathcal{G}^{n_{1}-2}\mathcal{G}'+\frac{b_{1}}{\mathcal{G}}\mathcal{G}')
\big\}+4r(2+6\gamma-r\gamma \eta')\big\{a_{1}
n_{1}(n_{1}-1)(n_{1}-2)\mathcal{G}^{n_{1}-3}\mathcal{G}'^{2}+a_{1}
n_{1}(n_{1}-1)\mathcal{G}^{n_{1}-2}\mathcal{G}''+\frac{b_{1}}{\mathcal{G}^{2}}
(\mathcal{G}\mathcal{G}''
\\\label{23}
&-&\mathcal{G}'^{2})\big\}\bigg\}\bigg],
\end{eqnarray}
where
\begin{eqnarray}\nonumber
\mathcal{G}&=&-\frac{2}{r^{2}e^{2\eta}}\big[(e^{\eta}-3)\xi'\eta'-(e^{\eta}-1)
(2\xi''+\xi'^{2})\big],
\\\nonumber
\mathcal{G}'&=&\frac{2}{r^{3}e^{2\eta}}\bigg[2(e^{\eta}-1)\xi'^{2}+(6-e^{\eta})
r\xi'\eta'^{2}
+2(e^{\eta}-1)(2\xi''-\xi''')+r\xi'{(e^{\eta}-3)\eta''-2(e^{\eta}-1)\xi''}
+\eta'\{2(3-e^{\eta})\xi'
\\\nonumber
&+&(3e^{\eta}-7)r\xi''+(e^{\eta}-2)r\xi'^{2}\}\bigg],
\\\nonumber
\mathcal{G}''&=&\frac{2}{r^{4}e^{2\eta}}\bigg[\xi'^{2}{6-6e^{\eta}
+(e^{\eta}-2)r^{2}\eta''}+(e^{\eta}-12)r^{2}\xi'\eta'^{3}-2\bigg\{\xi''
\big\{6(e^{\eta}-1)
-(2e^{\eta}-5)r^{2}\eta''+(e^{\eta}-1)r^{2}\xi''^{2}
\\\nonumber
&+&(e^{\eta}-1)r(2\xi'''-4\xi''')\big\}\bigg\}
+\eta'\bigg\{\xi'(6(e^{\eta}-3)+4(e^{\eta}-2)r^{2}\xi''-3(e^{\eta}-6)r^{2}\eta'')
-4(e^{\eta}-2)r\xi'^{2}+r(\xi'''(5e^{\eta}-11)r
\\\nonumber
&-&4(3e^{\eta}-7)\xi'')\bigg\}
-r\eta'^{2}{4(e^{\eta}-5)r\xi''-(e^{\eta}-6)\xi'+(e^{\eta}-4)r\xi'^{2}}
+r\xi'\big\{8(e^{\eta}-1)\xi''-4(e^{\eta}-3)\eta''+r\big((e^{\eta}-3)\eta'''
\\\nonumber
&-&2\xi'''(e^{\eta}-1)\big)\big\}\bigg].
\end{eqnarray}

\section*{Appendix B}
\renewcommand{\theequation}{B\arabic{equation}}
\setcounter{equation}{0}

The resulting field equations for model \textbf{2} turns out to be
\begin{eqnarray}\nonumber
\rho&=&\frac{e^{-2\eta}}{8r^{4}(1+\gamma)(1+2\gamma)}\bigg[8e^{\eta}
(e^{\eta}-1)r^2(1+2\gamma) -4r^{4}e^{2\eta}r^{4}(1+\gamma)(a_{2}
\mathcal{G}^{n_{2}}(b_{2} \mathcal{G}^{m}+1))+(a_{2}b_{2}
\mathcal{G}^{n_{2}+m-1}(n_{2}+m)
\\\nonumber
&+&a_{2}
n_{2}\mathcal{G}^{n_{2}-1})\bigg\{-16(e^{\eta}-1)^{2}\gamma+r^{2}
\bigg\{r^{2}(1+2\gamma)\xi'^{4}-2r^{2}(1+2\gamma)\xi'^{3}\eta'
-4\xi'\eta'\big\{2(e^{\eta}-3)(1+\gamma)+r^{2}(1+2\gamma)\xi''
\\\nonumber
&+&\xi'^{2}(8(\gamma+e^{\eta}(1+\gamma)))+r^{2}(1+2\gamma)(\eta'^{2}+4\xi'')\big\}
+4\big\{-2\gamma
\eta'^{2}+\xi''(4(e^{\eta}-1)(1+\gamma))\big\}+r^{2}(1+2\gamma)\xi''\bigg\}\bigg\}
\\\nonumber
&+&2r\bigg\{4(a_{2}b_{2}(n_{2}+m-1)(n_{2}+m)\mathcal{G}^{n_{2}+m-2}\mathcal{G}'+a_{2}
n_{2}(n_{2}-1)\mathcal{G}^{n_{2}-2}\mathcal{G}')\big\{-8(2+5\gamma)
+r(\eta'(10+27\gamma-2r\gamma \eta'))
\\\nonumber
&-&r(8+18\gamma+r(2+3\gamma)\eta')\xi''\big\}-8r(2+5\gamma)(1-2r\eta'+r^{2}\xi'')
\big\{a_{2}b_{2}(n_{2}+m-1)(n_{2}+m)(\mathcal{G}'^{2}(n_{2}+m-2)
\\\nonumber
&\times&\mathcal{G}^{n_{2}+m-3}+\mathcal{G}^{n_{2}+m-2}\mathcal{G}'')
+a_{2} n_{2}(n_{2}-1)(\mathcal{G}'^{2}(n_{2}-2)\mathcal{G}^{n_{2}-3}
+\mathcal{G}^{n_{2}-2}\mathcal{G}'')\big\}
+r^{2}\xi'^{2}\bigg\{re^{\eta}\gamma-2(8+18\gamma+r
\\\nonumber
&\times&(2+3\gamma)\eta')(a_{2}b_{2}(n_{2}+m-1)(n_{2}+m)
\mathcal{G}^{n_{2}+m-2}\mathcal{G}'+a_{2}
n_{2}(n_{2}-1)\mathcal{G}^{n_{2}-2}\mathcal{G}')-4r(2+5\gamma)\big\{a_{2}b_{2}(n_{2}+m-1)
\\\nonumber
&\times&(n_{2}+m)(\mathcal{G}'^{2}(n_{2}+m-2)\mathcal{G}^{n_{2}+m-3}
+\mathcal{G}^{n_{2}+m-2}\mathcal{G}'')+a_{2}
n_{2}(n_{2}-1)(\mathcal{G}'^{2}(n_{2}-2)\mathcal{G}^{n_{2}-3}
+\mathcal{G}^{n_{2}-2}\mathcal{G}'')\big\}\bigg\} +2e^{\eta}
\\\nonumber
&\times&\bigg\{16(2+5\gamma)(a_{2}b_{2}(n_{2}+m-1)(n_{2}+m)\mathcal{G}^{n_{2}+m-2}
\mathcal{G}'+a_{2}
n_{2}(n_{2}-1)\mathcal{G}^{n_{2}-2}\mathcal{G}')+2r\eta'\big\{r+2r\gamma+(2+3\gamma)
\\\nonumber
&\times&(a_{2}b_{2}(n_{2}+m-1)(n_{2}+m)\mathcal{G}^{n_{2}+m-2}
\mathcal{G}'+a_{2}
n_{2}(n_{2}-1)\mathcal{G}^{n_{2}-2}\mathcal{G}')\big\}+r^{3}\gamma
\xi''+4r(2+5\gamma)\big\{a_{2}b_{2}(n_{2}+m-1)
\\\nonumber
&\times&(n_{2}+m)(\mathcal{G}'^{2}(n_{2}+m-2)\mathcal{G}^{n_{2}+m-3}
+\mathcal{G}^{n_{2}+m-2}\mathcal{G}'')+a_{2}
n_{2}(n_{2}-1)(\mathcal{G}'^{2}(n_{2}-2)\mathcal{G}^{n_{2}-3}
+\mathcal{G}^{n_{2}-2}\mathcal{G}'')\big\}\bigg\}+r\xi'
\\\nonumber
&\times&\big\{2(-32
-74\gamma+r\eta'(2\gamma+r(2+3\gamma)\eta'))\big\}(a_{2}b_{2}(n_{2}+m-1)
(n_{2}+m)\mathcal{G}^{n_{2}+m-2}\mathcal{G}'+a_{2}
n_{2}(n_{2}-1)\mathcal{G}^{n_{2}-2}\mathcal{G}')
\\\nonumber
&-&e^{\eta}\gamma\big\{r(-4+r\eta')+4(a_{2}b_{2}(n_{2}+m-1)(n_{2}+m)
\mathcal{G}^{n_{2}+m-2}\mathcal{G}'+a_{2}
n_{2}(n_{2}-1)\mathcal{G}^{n_{2}-2}\mathcal{G}')\big\}+4r\bigg\{-2(4+9\gamma)
\\\nonumber
&+&r(2+5\gamma)\eta'\big\{a_{2}b_{2}(n_{2}+m-1)(n_{2}+m)(\mathcal{G}'^{2}(n_{2}+m-2)
\mathcal{G}^{n_{2}+m-3}+\mathcal{G}^{n_{2}+m-2}\mathcal{G}'')+a_{2}
n_{2}(n_{2}-1)
\\\label{21a}
&\times&
(\mathcal{G}'^{2}(n_{2}-2)\mathcal{G}^{n_{2}-3}+\mathcal{G}^{n_{2}-2}
\mathcal{G}'')\big\}\bigg\}\bigg\}\bigg],
\\\nonumber
p_{r}&=&\frac{e^{-2\eta}}{8r^{4}(1+\gamma)(1+2\gamma)}\bigg[-4e^{2\eta}r^{4}
(1+\gamma)(a_{2} \mathcal{G}^{n_{2}}(b_{2}
\mathcal{G}^{m}+1))+(a_{2}b_{2}
\mathcal{G}^{n_{2}+m-1}(n_{2}+m)+a_{2}
n_{2}\mathcal{G}^{n_{2}-1})\bigg\{-16\gamma
\\\nonumber
&\times&(e^{\eta}-1)^{2}+r^{2}\big\{r^{2}(1+2\gamma)\xi'^{4}-2r^{2}
(1+2\gamma)\xi'^{3}\eta'
-4\xi'\eta'(2(e^{\eta}-3)(1+\gamma)+r^{2}(1+2\gamma)\xi'')+\xi'^{2}
(8(e^{\eta}-1)
\\\nonumber
&\times&(1+\gamma)+r^{2}(1+2\gamma)(\eta'^{2}+4\xi''))+4\big\{2(1+\gamma)
\eta'^{2}+\xi''(4(e^{\eta}-1)(1+\gamma)+r^{2}(1+2\gamma)\xi'')\big\}\big\}\bigg\}
\\\nonumber
&+&2r(4e^{\eta}(e^{\eta}-1))r(1+2\gamma)+4(a_{2}b_{2}(n_{2}+m-1)(n_{2}+m)
\mathcal{G}^{n_{2}+m-2}\mathcal{G}'+a_{2}
n_{2}(n_{2}-1)\mathcal{G}^{n_{2}-2}\mathcal{G}')\bigg\{-8\gamma+r
\\\nonumber
&\times&\bigg\{-\eta'(4+\gamma+2r\gamma \eta')+r\gamma(-2+r
\eta'\xi'')+r^{2}\gamma
\xi'^{2}\bigg\{2(-2+r\eta')(a_{2}b_{2}(n_{2}+m-1)(n_{2}+m)
\mathcal{G}^{n_{2}+m-2}\mathcal{G}'
\\\nonumber
&+&a_{2}
n_{2}(n_{2}-1)\mathcal{G}^{n_{2}-2}\mathcal{G}')+r\big\{e^{\eta}
-4\big\{a_{2}b_{2}
(n_{2}+m-1)(n_{2}+m)(\mathcal{G}'^{2}(n_{2}+m-2)\mathcal{G}^{n_{2}+m-3}
+\mathcal{G}^{n_{2}+m-2}\mathcal{G}'')
\\\nonumber
&+&a_{2}
n_{2}(n_{2}-1)(\mathcal{G}'^{2}(n_{2}-2)\mathcal{G}^{n_{2}-3}
+\mathcal{G}^{n_{2}-2}\mathcal{G}'')
\big\}\big\}\bigg\}-8r\gamma(1-2r\eta'+r^{2}\xi'')
\big\{a_{2}b_{2}(n_{2}+m-1)(n_{2}+m) (\mathcal{G}'^{2}
\\\nonumber
&\times&(n_{2}+m-2)\mathcal{G}^{n_{2}+m-3}+\mathcal{G}^{n_{2}+m-2}
\mathcal{G}'')+a_{2}
n_{2}(n_{2}-1)(\mathcal{G}'^{2}(n_{2}-2)\mathcal{G}^{n_{2}-3}
+\mathcal{G}^{n_{2}-2}\mathcal{G}'')\big\}
+r\xi'\bigg\{-2(12+34\gamma
\\\nonumber
&+&r\eta'(8+14\gamma+4\gamma
\eta'))(a_{2}b_{2}(n_{2}+m-1)(n_{2}+m)\mathcal{G}^{n_{2}
+m-2}\mathcal{G}'+a_{2}
n_{2}(n_{2}-1)\mathcal{G}^{n_{2}-2}\mathcal{G}')
+e^{\eta}\big\{-r(4+4\gamma+r\gamma \eta')
\\\nonumber
&+&4(2+3\gamma)(a_{2}b_{2}(n_{2}+m-1)(n_{2}+m)\mathcal{G}^{n_{2}+m-2}\mathcal{G}'+a_{2}
n_{2}(n_{2}-1)\mathcal{G}^{n_{2}-2}\mathcal{G}')\big\}
+4r\gamma(-2+r\eta')\big\{a_{2}b_{2}(n_{2}+m-1)
\\\nonumber
&\times&(n_{2}+m)
(\mathcal{G}'^{2}(n_{2}+m-2)\mathcal{G}^{n_{2}+m-3}+\mathcal{G}^{n_{2}
+m-2}\mathcal{G}'')+a_{2}
n_{2}(n_{2}-1)(\mathcal{G}'^{2}(n_{2}-2)\mathcal{G}^{n_{2}-3}
+\mathcal{G}^{n_{2}-2}\mathcal{G}'')\big\}\bigg\}+2e^{\eta}
\\\nonumber&\times&
\bigg\{2(8\gamma+r(4+7\gamma)\eta')
(a_{2}b_{2}(n_{2}+m-1)(n_{2}+m)\mathcal{G}^{n_{2}+m-2}\mathcal{G}'+a_{2}
n_{2}(n_{2}-1)\mathcal{G}^{n_{2}-2}\mathcal{G}')+r\gamma
\big\{r^2\xi''+4\big\{a_{2}b_{2}
\\\nonumber
&\times&(n_{2}+m-1)(n_{2}+m)
(\mathcal{G}'^{2}(n_{2}+m-2)\mathcal{G}^{n_{2}+m-3}+\mathcal{G}^{n_{2}
+m-2}\mathcal{G}'')+a_{2}
n_{2}(n_{2}-1)(\mathcal{G}'^{2}(n_{2}-2)\mathcal{G}^{n_{2}-3}
\\\label{22a}
&+&\mathcal{G}^{n_{2}-2}\mathcal{G}'')\big\}\big\}\bigg\}\bigg\}\bigg\}\bigg],
\end{eqnarray}
\begin{eqnarray}\nonumber
p_{t}&=&\frac{e^{-2\eta}}{4r^{4}(1+\gamma)(1+2\gamma)}\bigg[-2e^{2\eta}r^{4}
(1+\gamma)(a_{2} \mathcal{G}^{n_{2}}(b_{2}
\mathcal{G}^{m}+1))+2(a_{2}b_{2}
\mathcal{G}^{n_{2}+m-1}(n_{2}+m)+a_{2}
n_{2}\mathcal{G}^{n_{2}-1})\bigg\{4(1+\gamma)
\\\nonumber
&\times&(e^{\eta}-1)^{2}+r^{2}
\big\{(-1+2e^{\eta}(1+\gamma))\xi'^{2}-2(-3+e^{\eta})(1+\gamma)\xi'\eta'
+\eta'^{2}+4(-1+e^{\eta})(1+\gamma)\xi''\big\}\bigg\}
-r\bigg\{4(a_{2}b_{2}
\\\nonumber
&\times&(n_{2}+m-1)(n_{2}+m)\mathcal{G}^{n_{2}+m-2}\mathcal{G}'+a_{2}
n_{2}(n_{2}-1)\mathcal{G}^{n_{2}-2}\mathcal{G}')\big\{8\gamma
+r(\eta'(-7\gamma-2r(1+\gamma)\eta')+r(2+6\gamma+r
\\\nonumber
&\times&(2+3\gamma)\eta')\xi'')\big\}
+8r\gamma(1-2r\eta'+r^{2}\xi'')\big\{(a_{2}b_{2}(n_{2}+m-1)(n_{2}+m)
(\mathcal{G}'^{2}(n_{2}+m-2)\mathcal{G}^{n_{2}+m-3}+\mathcal{G}^{n_{2}
+m-2}\mathcal{G}''))
\\\nonumber
&+&a_{2}
n_{2}(n_{2}-1)(\mathcal{G}'^{2}(n_{2}-2)\mathcal{G}^{n_{2}-3}
+\mathcal{G}^{n_{2}-2}\mathcal{G}'')\big\}+2e^{\eta}\bigg\{-16\gamma
(a_{2}b_{2}(n_{2}+m-1)(n_{2}+m)\mathcal{G}^{n_{2}+m-2}\mathcal{G}'+a_{2}
n_{2}
\\\nonumber
&\times&(n_{2}-1)\mathcal{G}^{n_{2}-2}\mathcal{G}')-r\eta'\big\{r+2r\gamma-2\gamma
(a_{2}b_{2}(n_{2}+m-1)(n_{2}+m)\mathcal{G}^{n_{2}+m-2}\mathcal{G}'+a_{2}
n_{2}(n_{2}-1)\mathcal{G}^{n_{2}-2}\mathcal{G}')\big\}
\\\nonumber
&+&r^{3}(1+\gamma)\xi''-4r\gamma
\big\{a_{2}b_{2}(n_{2}+m-1)(n_{2}+m)(\mathcal{G}'^{2}(n_{2}+m-2)
\mathcal{G}^{n_{2}+m-3}+\mathcal{G}^{n_{2}+m-2}\mathcal{G}'')
+a_{2}n_{2}(n_{2}-1)
\\\nonumber
&\times&(\mathcal{G}'^{2}(n_{2}-2)
\mathcal{G}^{n_{2}-3}+\mathcal{G}^{n_{2}-2}\mathcal{G}'')\big\}\bigg\}+
r^{2}\xi'^{2}\bigg\{e^{\eta}r(1+\gamma)
+2(2+6\gamma+r(2+3\gamma)\eta')(a_{2}b_{2}(n_{2}+m-1)(n_{2}+m)
\\\nonumber
&\times& \mathcal{G}^{n_{2}+m-2}\mathcal{G}'+a_{2}
n_{2}(n_{2}-1)\mathcal{G}^{n_{2}-2}\mathcal{G}')+4r\gamma
\bigg\{a_{2}b_{2}(n_{2}+m-1)(n_{2}+m)(\mathcal{G}'^{2}(n_{2}+m-2)
\mathcal{G}^{n_{2}+m-3}+\mathcal{G}^{n_{2}+m-2}\mathcal{G}'')
\\\nonumber
&+&a_{2}
n_{2}(n_{2}-1)(\mathcal{G}'^{2}(n_{2}-2)\mathcal{G}^{n_{2}-3}+\mathcal{G}^{n_{2}-2}\mathcal{G}'')
\bigg\}\bigg\}+r\xi'-2(-10\gamma+r\eta'(2+6\gamma+r(2+3\gamma)\eta')
)(a_{2}\eta_{2}(n_{2}+m-1)
\\\nonumber
&\times&(n_{2}+m)\mathcal{G}^{n_{2}+m-2}\mathcal{G}'+a_{2}
n_{2}(n_{2}-1)\mathcal{G}^{n_{2}-2}\mathcal{G}')+e^{\eta}\big(-r(-2+r(1+\gamma)\eta')+4\gamma
(a_{2}\eta_{2}(n_{2}+m-1)(n_{2}+m)
\\\nonumber
&\times& \mathcal{G}^{n_{2}+m-2}\mathcal{G}'+a_{2}
n_{2}(n_{2}-1)\mathcal{G}^{n_{2}-2}\mathcal{G}')\big)+4r(2+6\gamma-r\gamma
\eta')\big\{a_{2}b_{2}(n_{2}+m-1)(n_{2}+m)(\mathcal{G}'^{2}(n_{2}+m-2)\mathcal{G}^{n_{2}+m-3}
\\\label{23a}
&+&\mathcal{G}^{n_{2}+m-2}\mathcal{G}'')+a_{2}
n_{2}(n_{2}-1)(\mathcal{G}'^{2}(n_{2}-2)\mathcal{G}^{n_{2}-3}+\mathcal{G}^{n_{2}-2}
\mathcal{G}'')\big\}\bigg\}\bigg].
\end{eqnarray}

\section*{Appendix C}
\renewcommand{\theequation}{C\arabic{equation}}
\setcounter{equation}{0}

The field equations for model \textbf{3} are
\begin{eqnarray}\nonumber
\rho&=&\frac{e^{-2\eta}}{8r^{4}(1+\gamma)(1+2\gamma)}\bigg[8e^{\eta}(e^{\eta}-1)
r^2(1+2\gamma)
-4r^{4}e^{2\eta}r^{4}(1+\gamma)(a_3\mathcal{G}^{n_3}+b_3)(a_4\mathcal{G}^{n_3}
+b_4)^{-1}+(a_2\mathcal{G}^{n_3}+b_4)^{-2}
\\\nonumber
&\times&n_3(b_4a_1-b_3a_2)\mathcal{G}^{n_3-1}\bigg\{-16(e^{\eta}-1)^{2}\gamma+r^{2}
\big\{r^{2}(1+2\gamma)\xi'^{4}
-2r^{2}(1+2\gamma)\xi'^{3}\eta'-4\xi'\eta'\big\{2(e^{\eta}-3)(1+\gamma)
\\\nonumber
&+&r^{2}(1+2\gamma)\xi''+\xi'^{2}(8(\gamma+e^{\eta}(1+\gamma)))+r^{2}(1+2\gamma)
(\eta'^{2}+4\xi'')\big\}+4\big\{-2\gamma
\eta'^{2}+\xi''(4(e^{\eta}-1)(1+\gamma))\big\}
\\\nonumber
&+&r^{2}(1+2\gamma)\xi''\big\}\bigg\}
+2r\bigg\{4\mathcal{A}\big\{-8(2+5\gamma)+r(\eta'(10+27\gamma-2r\gamma
\eta'))-r(8+18\gamma+r(2+3\gamma)\eta')\xi''\big\}-8r
\\\nonumber
&\times&(2+5\gamma)(1-2r\eta'+r^{2}\xi'')\mathcal{B}
+r^{2}\xi'^{2}\big\{re^{\eta}\gamma-2(8+18\gamma+r(2+3\gamma)\eta')\mathcal{A}
-4r(2+5\gamma)\mathcal{B}\big\}+2e^{\eta}\big\{16(2+5\gamma)\mathcal{A}
\\\nonumber
&+&2r\eta'(r+2r\gamma+(2+3\gamma)\mathcal{A})+r^{3}\gamma
\xi''+4r(2+5\gamma)\mathcal{B}\big\}+r\xi'\big\{2(-32-74\gamma
+r\eta'(2\gamma+r(2+3\gamma)\eta'))\big\}\mathcal{A}
\\\label{21c}
&-&e^{\eta}\gamma(r(-4+r\eta')+4\mathcal{A})
+4r\big\{-2(4+9\gamma)+r(2+5\gamma)\eta'\mathcal{B}\big\}\bigg\}\bigg],
\\\nonumber
p_{r}&=&\frac{e^{-2\eta}}{8r^{4}(1+\gamma)(1+2\gamma)}\bigg[-4e^{2\eta}r^{4}
(1+\gamma) (a_3\mathcal{G}^{n_3}+b_3)(a_4\mathcal{G}^{n_3}+b_4)^{-1}
+n_3(b_4a_1-b_3a_2)\mathcal{G}^{n_3-1}(a_2\mathcal{G}^{n_3}+b_4)^{-2}
\\\nonumber
&\times&\bigg\{-16(e^{\eta}-1)^{2}\gamma+r^{2}
\big\{r^{2}(1+2\gamma)\xi'^{4}-2r^{2}(1+2\gamma)\xi'^{3}\eta'-4\xi'\eta'
(2(e^{\eta}-3)(1+\gamma) +r^{2}(1+2\gamma)\xi'')
\\\nonumber
&+&\xi'^{2}\big\{8(e^{\eta}-1)(1+\gamma)+r^{2}(1+2\gamma)(\eta'^{2}+4\xi'')\big\}
+4\big\{2(1+\gamma)\eta'^{2}+\xi''(4(e^{\eta}-1)(1+\gamma)+r^{2}(1+2\gamma)\xi'')
\big\}\big\}\bigg\}
\\\nonumber
&+&2r(4e^{\eta}(e^{\eta}-1))r(1+2\gamma)+4\mathcal{A}\bigg[-8\gamma+r\bigg\{-\eta'
(4+\gamma+2r\gamma \eta')+r\gamma(-2+r \eta'\xi'')+r^{2}\gamma
\xi'^{2}(2(-2+r\eta')\mathcal{A}
\\\nonumber
&+&r(e^{\eta}-4\mathcal{B}))-8r\gamma(1-2r\eta'+r^{2}\xi'')\mathcal{B}
+r\xi'\big\{-2\big(12+34\gamma+r\eta'(8+14\gamma+4\gamma
\eta')\big)\mathcal{A}+e^{\eta}(-r(4+4\gamma+r\gamma \eta')
\\\label{22c}
&+&4(2+3\gamma)\mathcal{A})+4r\gamma(-2+r\eta')\mathcal{B}\big\}+2e^{\eta}
\big\{2(8\gamma+r(4+7\gamma)\eta')\mathcal{A}+r\gamma(r^2\xi''+4\mathcal{B})
\big\}\bigg\}\bigg]\bigg],
\\\nonumber
p_{t}&=&\frac{e^{-2\eta}}{4r^{4}(1+\gamma)(1+2\gamma)}\bigg[-2e^{2\eta}r^{4}(1+\gamma)
(a_3\mathcal{G}^{n_3}+b_3)(a_4\mathcal{G}^{n_3}+b_4)^{-1}
+2n_3(b_4a_1-b_3a_2)\mathcal{G}^{n_3-1}(a_2\mathcal{G}^{n_3}+b_4)^{-2}
\\\nonumber
&\times&\bigg\{4(e^{\eta}-1)^{2}(1+\gamma)+r^{2}
\big\{(-1+2e^{\eta}(1+\gamma))\xi'^{2}-2(-3+e^{\eta})(1+\gamma)\xi'\eta'+\eta'^{2}
+4(-1+e^{\eta})(1+\gamma)\xi''\big\}\bigg\}
\\\nonumber
&-&r\bigg\{4f_{\mathcal{G}}'\big\{8\gamma+r(\eta'(-7\gamma-2r(1+\gamma)\eta')
+r(2+6\gamma+r(2+3\gamma)\eta')\xi'')\big\}+8r\gamma(1-2r\eta'+r^{2}\xi'')\mathcal{B}
+2e^{\eta}
\\\nonumber
&+&\big\{-16\gamma \mathcal{A}-r\eta'(r+2r\gamma-2\gamma
\mathcal{A})+r^{3}(1+\gamma)\xi''-4r\gamma
\mathcal{B}\big\}+r^{2}\xi'^{2}\big\{e^{\eta}r(1+\gamma)+2(2+6\gamma+r(2+3\gamma)
\eta')\mathcal{A}
\\\nonumber
&+&4r\gamma
\mathcal{B}\big\}+r\xi'-2(-10\gamma+r\eta'(2+6\gamma+r(2+3\gamma)\eta')
)\mathcal{A}+e^{\eta}\big(-r(-2+r(1+\gamma)\eta')+4\gamma
\mathcal{A}\big)
\\\label{23c}
&+&4r(2+6\gamma-r\gamma \eta')\mathcal{B}\bigg\}\bigg],
\end{eqnarray}
where
\begin{eqnarray}\nonumber
\mathcal{A}&=&\frac{\mathcal{G}'}{(
a_{4}\mathcal{G}^{n_{3}}+b_{4})^{3}}\bigg[
n_{3}\bigg\{(\mathcal{G}^{2n_{3}-2}b_{3}
a_{4}^{2}n_{3}-\mathcal{G}^{2n_{3}-2}b_{4}a_{3} a_{4}n_{3}
+\mathcal{G}^{2n_{3}-2}b_{3}a_{4}^{2}-\mathcal{G}
^{2n_{3}-2}b_{4}a_{3}a_{4}-\mathcal{G}^{n_{3}-2}b_{3}b_{4}
a_{4}n_{3}
\\\nonumber
&+&\mathcal{G}^{n_{3}-2}b_{4}^{2}a_{3}n_{3}+\mathcal{G}^{n_{3}-2}
b_{3}b_{4}a_{4}-\mathcal{G}^{n_{3}-2}b_{4}^{2}a_{3}\bigg\}\bigg],
\\\nonumber
\mathcal{B}&=&\mathcal{G}''\mathcal{A}+\frac{\mathcal{G}'^{2}}{(a_{4}\mathcal{G}^n_{3}+b_{4}
)^{4}}\bigg[-n_{3}\bigg\{
\mathcal{G}^{n-3}b_{3}b_{4}^{2}a_{4}n_{3}^{2}-
\mathcal{G}^{n-3}b_{4}^{3}a_{3}n_{3}^{2}+\mathcal{G}^{3n-3}b_{3}a_{4}^{3}
n_{3}^{2}-\mathcal{G}^{3n-3}b_{4}a_{3}a_{4}^{2}n_{3}^{2}-4
\mathcal{G}^{2n-3}b_{3}b_{4}a_{4}^{2}n_{3}^{2}
\\\nonumber
&+&4\mathcal{G}^{2n-3
}b_{4}^{2}a_{3}a_{4}n_{3}^{2}-3\mathcal{G}^{n-3}b_{3}b_{4}^{2}
a_{4}n_{3}+3\mathcal{G}^{n-3}b_{4}^{3}a_{3}n+3
a_{4}^{3}n_{3}\mathcal{G}^{3n-3}b_{3}-3\mathcal{G}^{3n-3}b_{4}a_{3}
a_{4}^{2}n_{3}+2\mathcal{G}^{n-3}b_{3}b_{4}^{2}a_{4}
\\\nonumber
&-&2\mathcal{G}^{n-
3}b_{4}^{3}a_{3}+2\mathcal{G}^{3n-3}b_{3}a_{4}^{3}-2\mathcal{G}
^{3n-3}b_{4}a_{3}a_{4}^{2}+4\mathcal{G}^{2n-3}b_{3}b_{4}
a_{4}^{2}-4\mathcal{G}^{2n-3}b_{4}^{2}a_{3}a_{4}\bigg\}\bigg].
\end{eqnarray}
\\\\
\textbf{Data Availability Statement:} This manuscript has no
associated data.

\end{document}